\documentclass[preprint]{aastex}

\shorttitle{Disk masses of the GV Tau Binary}
\shortauthors{Sheehan and Eisner}

\begin{document}

\title{Constraining the Disk Masses of the Class I Binary Protostar GV Tau}
\author{Patrick D. Sheehan and Josh A. Eisner}
\affil{Steward Observatory, University of Arizona}
\affil{933 N. Cherry Avenue, Tucson, AZ, 85721}
\email{psheehan@email.arizona.edu}

\begin{abstract}

We present new spatially resolved 1.3 mm imaging with CARMA of the GV Tau system. GV Tau is a Class I binary protostar system in the Taurus Molecular Cloud, the components of which are separated by 1.2". Each protostar is surrounded by a protoplanetary disk, and the pair may be surrounded by a circumbinary envelope. We analyze the data using detailed radiative transfer modeling of the system. We create synthetic protostar model spectra, images, and visibilities and compare them with CARMA 1.3 mm visibilities, an HST near-infrared scattered light image, and broadband SEDs from the literature to study the disk masses and geometries of the GV Tau disks. We show that the protoplanetary disks around GV Tau fall near the lower end of estimates of the Minimum Mass Solar Nebula, and may have just enough mass to form giant planets. When added to the sample of Class I protostars from \citet{Eisner2012} we confirm that Class I protostars are on average more massive than their Class II counterparts. This suggests that substantial dust grain processing occurs between the Class I and Class II stages, and may help to explain why the Class II protostars do not appear to have, on average, enough mass in their disks to form giant planets.

\end{abstract}


\section{Introduction}

The process of star formation begins with a roughly spherical cloud of gas and dust in hydrostatic equilibrium that has yet to begin collapsing under the force of gravity to form a protostar. As the collapse proceeds, conservation of angular momentum forces most of the in-falling material to form a disk rather than accrete directly onto the forming protostar. Viscosity in the disk then transports mass inwards and angular momentum outward, allowing matter to accrete from the disk onto the central protostar. Eventually the material from the in-falling spherical envelope is depleted onto the massive protostellar disk. In turn, the material from the disk is then deposited onto the pre-main sequence star until the disk is tenuous and the central star is exposed. At the same time dust grains in the disk coagulate to form larger and larger bodies, which eventually may grow into planets.

Young stars are typically classified according to the shape of their spectral energy distributions (SEDs) \citep[e.g.][]{Lada1987,Andre1993} and their bolometric temperatures \citep[e.g.][]{Enoch2009}. Class 0 Young Stellar Objects (YSOs) have SEDs that are highly obscured at optical and near to mid infrared wavelengths and peak at far-infrared or sub-millimeter wavelengths, corresponding to bolometric temperatures below 100 K. These objects are believed to be young stars that are still enveloped by their natal envelopes. There is evidence suggesting that a few Class 0 protostars are surrounded by rotationally supported protostellar disks \citep[e.g.][]{Tobin2013}. However, it is not yet clear that Class 0 YSOs in general possess disks. Class I YSO SED's rise steeply in the near-infrared, peak in the mid-infrared, and often have significant obscuration of their central protostars. They are also characterized by bolometric temperatures below about 600 K. This class likely represents protostars surrounded by massive disks still embedded in their original envelopes. Class II YSOs have SEDs that are flatter at near-infrared wavelengths and show some light from the central protostar. They are thought to represent pre-main sequence stars encompassed by massive protoplanetry disks. Class III YSO SEDs are dominated by the light from the central protostar, and have little or no infrared excess arising from an optically thin disk of matter.

The mass of the circumstellar disk at each of these stages is an important indicator for the evolution of circumstellar mass during star and planet formation. Disks that are too massive may be subject to gravitational instabilities that could help to grow protostellar mass quickly. Gravitational instabilities leading to rapid mass accretion may help to rectify the discrepancy between observed envelope-to-disk and disk-to-star mass accretion rates \citep[e.g.][]{Kenyon1987}. Conversely, disks with too little mass may not have enough material to form giant planets \citep[e.g.][]{Weidenschilling1977, Desch2007}.

The masses of protostellar disks and envelopes are usually measured from millimeter wavelength observations. If the matter is optically thin, as is much of the circumstellar material around protostars, then the millimeter flux is proportional to the dust mass. In order, however, to make the conversion between millimeter flux and total mass it is necessary to know the temperature distribution throughout the disk, the opacity of the dust in the disk, as well as the gas-to-dust mass ratio. Furthermore, dense regions in the disk can be optically thick and hide material from sight. For Class I objects, which are surrounded by both a disk and its natal envelope, disentangling the disk and envelope masses is also a challenge. The best method for overcoming these difficulties and unambiguously determining the mass of the protostellar disk is through detailed radiative transfer modeling of resolved imaging.

Studies using radiative transfer modeling to match SEDs have historically been used to place constraints on the distribution of matter around young stars \citep[e.g.][]{Adams1987, Kenyon1993, Robitaille2007}. Such modeling, however, can be subject to significant degeneracies. For example, it is difficult to use a SED to distinguish between a flattened disk-like envelope \citep{Ulrich1976, Terebey1984} and flared edge-on disks \citep{Chiang1999}. To break these degeneracies, additional imaging datasets such as short wavelength scattered light images or millimeter continuum images can be modeled to provide new constraints on circumstellar structure. Modeling of multiple datasets has previously been used to determine the circumstellar mass distribution of young stars more accurately than was possible by modeling a single dataset by itself \citep[e.g.][]{Osorio2003, Wolf2003, Eisner2005, Eisner2012}.

Disk masses for Class 0 protostars (ages $\lesssim 0.2$Myr), if indeed disks are present, have been suggested to be high \citep[$\gtrsim 0.05-0.1$ M$_{\odot}$;][]{Jorgensen2009}. Mass accretion rates of Class 0 protostars have also been estimated to be high ($\gtrsim 10^{-5}$ M$_{\odot}$ yr$^{-1}$) from SED fitting \citep[e.g.][]{Jayawardhana2001}, outflow measurements \citep[e.g.][]{Bontemps1996}, and lifetime measurements from statistical arguments \citep[e.g.][]{Andre1994,Barsony1994}. These high disk masses and accretion rates suggest that the disks around these protostars may be gravitationally unstable. Conversely, the masses of Class II disks in Taurus and Orion (ages $\sim 1-5$Myr) have been well studied and are found to have a median mass of 0.001 M$_{\odot}$, with $\lesssim$10\% of systems having disk masses higher than 0.01 M$_{\odot}$ and $\lesssim$1\% with disk masses greater than 0.1 M$_{\odot}$ \citep[e.g.][]{Eisner2008, Andrews2013}. These median masses are low compared with the amount of matter needed to form giant planets, estimated to be 0.01 - 0.1 M$_{\odot}$ \citep[e.g.][]{Weidenschilling1977,Desch2007}. The millimeter wavelength observations used to make these measurements, however, are only sensitive to particles smaller than $\sim 1$mm. It might be the case that significant dust processing and grain growth has already occurred in these systems, effectively hiding the mass in the disk in larger undetectable bodies.

Class I YSOs thus may represent a transitional stage between massive, highly unstable protoplanetary disks to stable disks in which planet formation is progressing. The disks around Class I YSOs may also more accurately represent the initial mass budget of disks for forming planets as they are younger and presumably grain growth is less advanced. 

Previous radiative transfer modeling studies of the masses of Class I disks in Taurus and Ophiuchus (ages $\sim 0.2-0.5$ Myr) using millimeter continuum images \citep[e.g.][]{Jorgensen2009} or SEDs, scattered light images, and millimeter images \citep{Osorio2003, Wolf2003, Eisner2005, Eisner2012} find disk masses ranging from 0.005-1 M$_{\odot}$. \citet{Eisner2012} finds a median disk mass for their sample of 0.01 M$_{\odot}$. They also find, however, that the mass within 100 AU, where planets form, has a median of 0.008 M$_{\odot}$. If the mass measured using millimeter emission traces the entire disk mass, there is likely not enough matter for forming giant planets, which may require as much as 0.1 M$_{\odot}$ \citep{Eisner2012}.

Binary stars are particularly interesting candidates for disk mass studies, not only because they allow measurements of two disk masses simultaneously, but also because a significant fraction of young stars are formed with companions \citep[e.g.][]{Abt1976,Raghavan2010}, so their properties are important for understanding the evolution of disk masses and planet formation for a large portion of young stars. Furthermore, disks in young binary systems are coeval. Similarities and differences in the properties of each individual system will therefore highlight nuances in the progression of star and planet formation.

In this paper we study the Class I binary GV Tau. GV Tau (IRAS 04263-2426, Haro 6-10) is located in the Taurus Molecular Cloud Complex, at a distance of 140 pc \citep{Mamajek2008}. GV Tau was first discovered to be a binary by \citet{Leinert1989} using speckle interferometry, and has since been resolved at near-infrared, millimeter, and centimeter wavelengths \citep{Koresko1999, Reipurth2004, Roccatagliata2011, Guilloteau2011}. The binary consists of a bright optical source, GV Tau S, and its companion, GV Tau N, located 1.2" north of its southern counterpart. At the distance of Taurus this projected separation corresponds to 170 AU. GV Tau N is 100 times fainter than GV Tau S at optical wavelengths but becomes bright in the near- and mid-infrared \citep{Leinert1989, Koresko1999, Roccatagliata2011}. \citet{Doppmann2008} find that the GV Tau N and S have stellar masses of 0.8 and 0.5 M$_{\odot}$ and temperatures of 3800 and 4100 K respectively.

Both components of the GV Tau binary have been found to be highly variable in the near-infrared on timescales as short as a month \citep{Leinert2001}. \citet{Leinert2001} attribute the variability of GV Tau N to variable accretion and suggest that the variability of GV Tau S is due to inhomogeneities in its accretion disk. \citet{Doppmann2008} find that GV Tau S has a variable radial velocity and suggest that GV Tau S may be a multiple system with a companion with mass M$_{\star} < 0.15$ M$_{\odot}$ (mass ratio $>3$) and $a < 0.35$ AU.

A number of previous studies have attempted to constrain the distribution of material around each component of the GV Tau binary. Early near-infrared imaging studies by \citet{Menard1993} suggested that the binary pre-main sequence stars were surrounded by a flattened circumbinary envelope or disk, and potentially circumstellar disks around each component. More recent studies have suggested that GV Tau N is surrounded by an edge on disk while GV Tau S's disk is close to face on, and that both components are surrounded by a common envelope, the composition of which is similar to that of the interstellar medium \citep{Roccatagliata2011}. \citet{Guilloteau2011} modeled Plateau de Bure Interferometer 1.3mm visibilities for the GV Tau binary and found the disks to be optically thick with radii around 15 AU.

In this work we use detailed radiative transfer modeling of new CARMA 1.3mm visibilities along with HST scattered light imaging and broadband SEDs to expand on previous works and more accurately constrain the structure and properties of the GV Tau binary young stellar objects.

\section{Observations \& Data Reduction}

\subsection{CARMA Observations \& Data Reduction}

\begin{figure}[h!]
\centering
\includegraphics[width=3in]{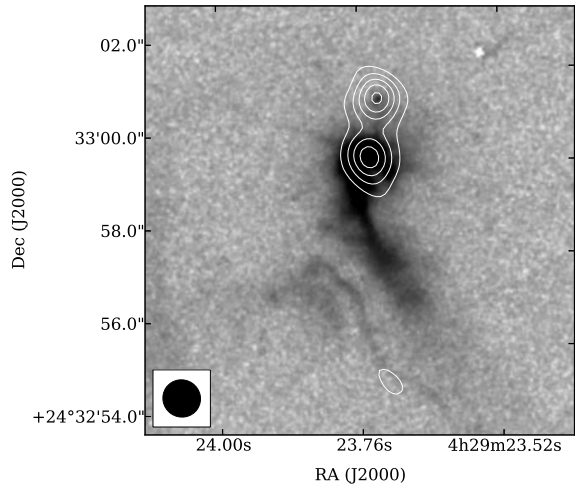}
\caption{0.8 $\mu$m HST scattered light image of GV Tau in grayscale with the 1.3 millimeter CARMA image overplotted as contours. Both images have high enough spatial resoultion and sensitivity to resolve the binary. The beam size of the millimeter image is shown in the bottom left.}
\label{implot}
\end{figure}

We observed GV Tau on 2010 October 29 with the Combined Array for Research in Millimeter-wave Astronomy (CARMA). Our observations were taken in CARMA's C configuration, with baselines ranging from 20 - 350 m, corresponding to an angular resolution of 1" and a largest resolvable scale ($\theta_{MRS} \sim 0.5 \lambda / B_{min}$) of 6.5". The CARMA correlator was in wideband mode, with a local oscillator (LO) frequency of 227 GHz, an intermediate frequency (IF) band $\pm 1-9$ GHz from the LO. Eight 500 MHz bands were placed evenly spaced in each of the sidebands, for a total continuum bandwidth of 8 GHz. Our observations were taken during the same track as two other young stars in Taurus with cycles of 19 minutes consisting of 5 minute integrations for each science target and 4 minutes for our gain calibrator, 3C111. We also observed the quasar 3C84 at the beginning of the track for bandpass calibration. The total on-source integration time for GV Tau was 50 minutes, and the total length of the track was 3 hours and 15 minutes.

The calibration of our data was done using the CASA and \emph{MIRIAD} data reduction packages. We applied a series of calibration corrections to the data, beginning with a correction for instrumental phase drifts from differences in line lengths. Next, we used 3C84 to estimate the bandpass responses and correct for variations in flux across the channels of each band. Antenna 8 was used as the reference antenna throughout the calibration process. After applying the bandpass corrections to the data we used the CASA  gain calibration routine on 3C111 to determine the time dependent gain corrections and interpolated to apply them to the data. No flux calibrator was observed during this track, so we scaled the visibilities using a flux of 1.94~Jy for 3C111 at 1 mm as measured by the SMA on October 28, 2010\footnotemark[1].

\footnotetext[1]{Can be found at http://sma1.sma.hawaii.edu/callist/callist.html}

\begin{figure}[h!]
\centering
\includegraphics[width=3in]{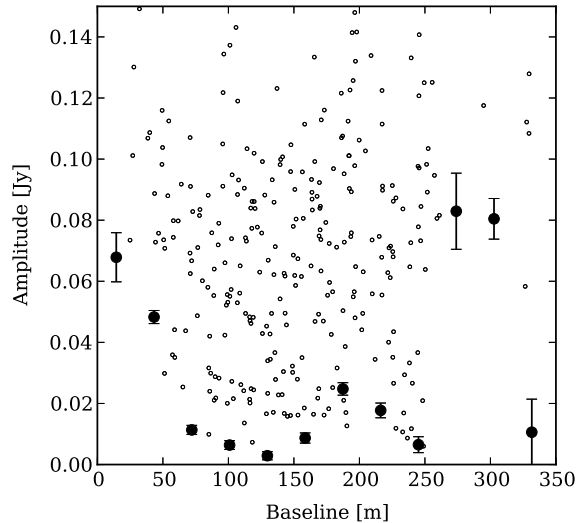}
\caption{1.3 mm visibilities for the GV Tau system. We plot the amplitudes of the one dimensional azimuthally averaged visbilities with solid points, while the open circles show the the amplitudes of the visibilities averaged using a two-dimensional grid. We average the data coherently, so phase noise in the data may result in average amplitudes which are lower than the amplitudes for the unaveraged data.}
\label{visplot}
\end{figure}

\begin{figure*}[h!]
\centering
\includegraphics[width=6in]{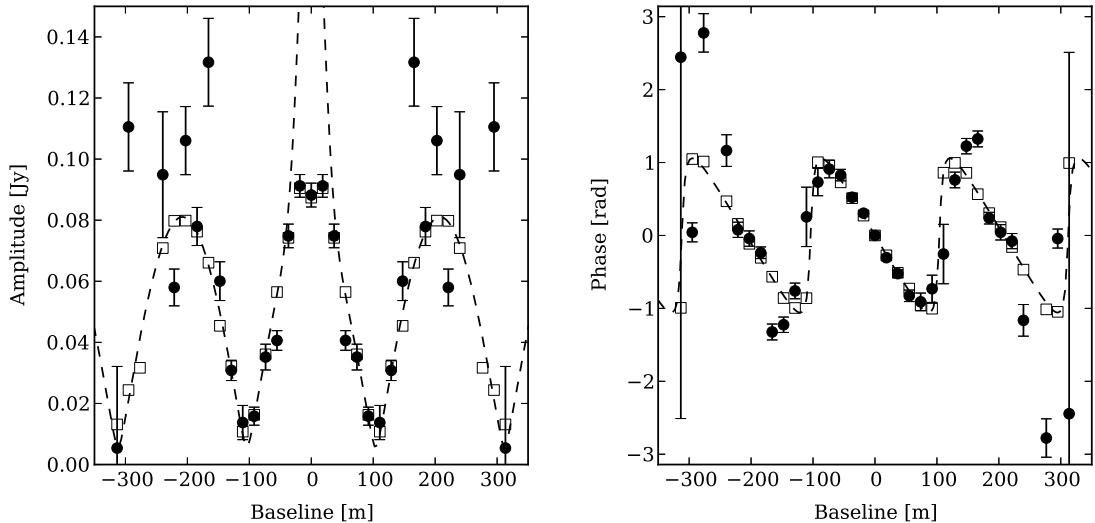}
\caption{1.3 mm visibilities, shown as filled circles, plotted and averaged along the axis of the binary. On the left we show the amplitude and on the right we show the phase of the complex visibilities. The data are perfectly symmetric across the zero-baseline line because the complex visibilities are Hermitian. We have over plotted the best fit double point source plus gaussian model as open squares and a dashed line. The open squares represent the best fit model sampled at the same binned $uv$ points as our data, while the dashed line shows the best fit model if the $uv$ plane were perfectly sampled.}
\label{visplot_model}
\end{figure*}

After calibrating the visibilities, we Fourier transformed our data to obtain an image, and we CLEANed the resulting image to deconvolve the image and the dirty beam. The imaging provides a nice visualization of the system, however we perform most of the analysis for this paper in the visibility plane, where we do not have to contend with beam effects. We plot the millimeter contours in Figure \ref{implot} and the visibilities in Figure \ref{visplot}. To better demonstrate that the target is a binary using the visibilities we plot the visibilities averaged in bins along a baseline parallel to the binary in Figure \ref{visplot_model}.

To model our targets individually we needed to separate the contribution to the measured visibilities of each member of the binary. To do this we fit the combined visibilities with both a double point source model, leaving the centroids and fluxes as free parameters, and a double two dimensional gaussian model, with the widths, centroids, fluxes, inclination, and position angle as free parameters. From our best fit we find that both components are unresolved, or at best marginally resolved, by our observations. 

We also find that our double point source models underpredict the flux of our targets at short baselines. We can improve the model fit to the data by adding a gaussian source with a large spatial extent to the model, likely representing large scale circumbinary structure. This gaussian has a FWHM of $\sim 5$", making it significantly larger than the binary. Given our limited coverage of short baselines, however, the outer scale of this structure is difficult to constrain. The best fit model is plotted over our data in Figure \ref{visplot_model}. To determine the visibilities for a single component of the binary we first remove the component arising from the large scale circumbinary material. We then subtract the best fit model for the other component from the visibilities. We do this for both the best fit point source and gaussian models and find that the difference in the resulting single component visibilities is negligible.

We binned the visibilities into both a two dimensional grid as well as an annular grid to increase the signal to noise ratio for our data. We gridded the visibilities with a weighted average of real and imaginary components of the visibilities within each grid cell, with weights determined by our calibration.

\subsection{Scattered Light Imaging}

\begin{figure}[h!]
\centering
\includegraphics[width=3in]{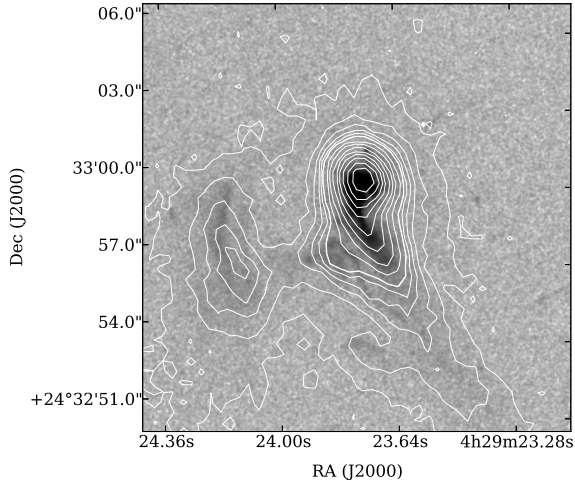}
\caption{HST 0.8 $\mu$m image of GV Tau in grayscale with contours of the SDSS $i$-band image overplotted. We matched the SDSS image to the HST image based on features in the scattered light image in order to transfer astrometry from the SDSS image to the HST image. This figure shows our best match, which we used for the transfer. This figure demonstrates that the HST image has much higher spatial resolution than the SDSS image, so we use the HST image for our modeling.}
\label{astrplot}
\end{figure}

We downloaded an archival \emph{Hubble Space Telescope} Widefield and Planetary Camera 3 (WFPC3) near-infrared 0.8 $\mu$m scattered light image of GV Tau from the Hubble Legacy Archive (HST Program 7387, PI: Stapelfeldt). The image was previously calibrated, with the exception of cosmic ray removal so we removed the cosmic ray hits from the image using the COSMICS program \citep{vanDokkum2001}. Finally, we scaled the data to units of ergs cm$^{-2}$ s$^{-1}$ $\AA^{-1}$ using the appropriate scaling factor from the fits header. We calculated uncertainties for the image from the square root of the counts frame of the data multiplied by the scaling factor to convert the image to a real flux value.

The HST image of GV Tau lacks background stars to be used to for determining astrometry of the image. Instead we used a widefield Sloan Digital Sky Survey (SDSS), Data Release 10, 0.75 $\mu$m scattered light image of GV Tau which does have background stars to determine the astrometry and transfer it to the HST images. We used the HST image in our modeling rather than the SDSS image because the HST image has higher resolution and shows significantly more structure than the SDSS image. We used SExtractor \citep{Bertin1996} and SCAMP \citep{Bertin2006} to locate point sources in the SDSS images and find an astrometric solution for the image. We then used distinctive features of the scattered light surrounding the southern component to align the HST and SDSS images and transfer the astrometry to the HST image. Figure \ref{astrplot} shows a plot of our alignment of the SDSS and HST images, and Figure \ref{implot} shows the HST image with the millimeter contours over-plotted. The uncertainty in the SDSS image astrometry is 0.2", and we estimate that the uncertainty in the HST image is 0.3".

\subsection{Photometric Data from the Literature}

\begin{figure*}[h!]
\centering
\includegraphics[width=6in]{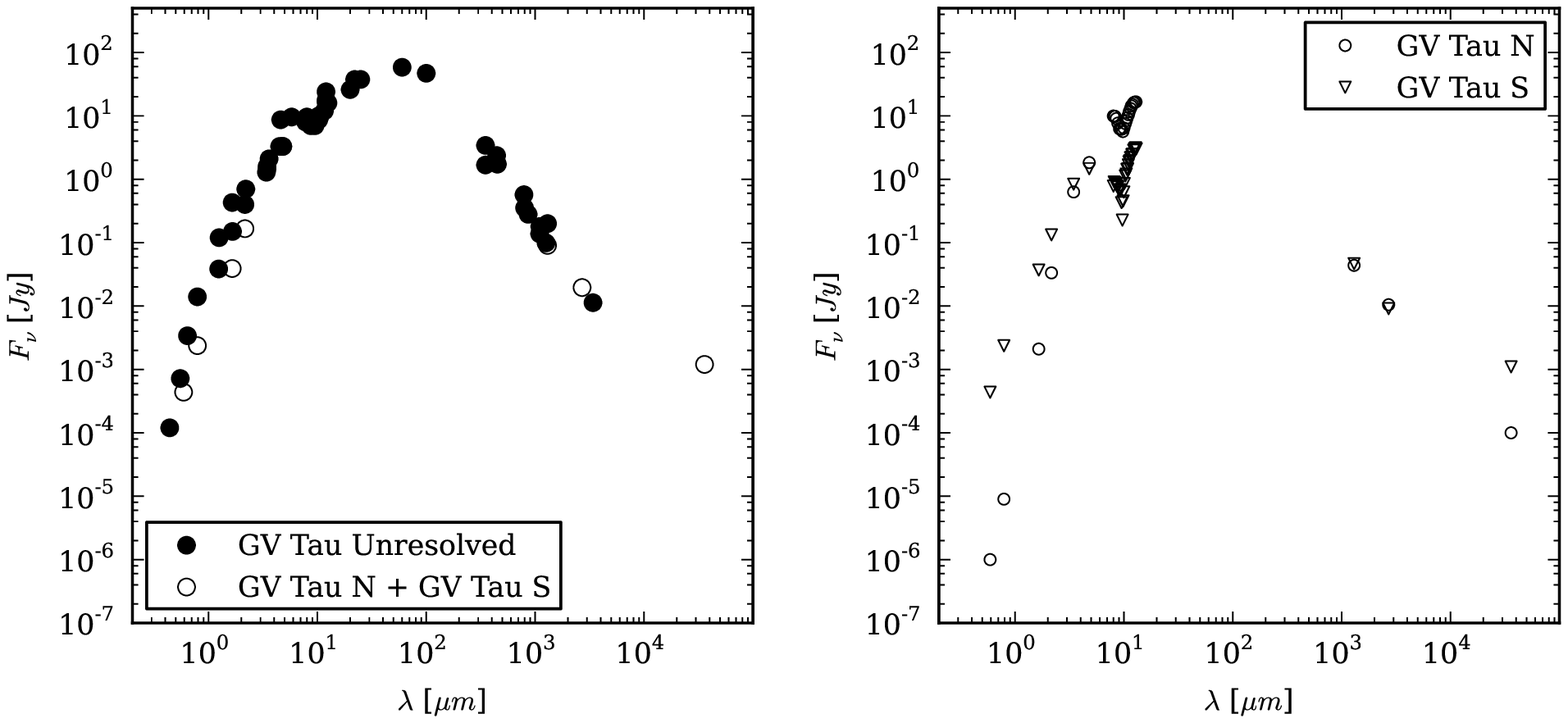}
\caption{SEDs for GV Tau using data from the literature. We plot the unresolved photometry on the left as filled circles as well as the sums of the resolved photometry as open circles. At most wavelengths, these lie on top of each other. The sums of the resolved near-infrared photometry likely fall below the unresolved photometry due to the smaller aperture used for the unresolved photometry. The unresolved photometry is likely more sensitive to the extended scattered light structure. In the right panel we plot the resolved photometry for both components of the GV Tau system.}
\label{specplot}
\end{figure*}

\begin{deluxetable}{ccc}
\tablecaption{Unresolved photometry of GV Tau}
\tablewidth{0pt}
\tabletypesize{\tiny}
\tablehead{
\colhead{$\lambda$} &  \colhead{F$_{\nu}$} & \colhead{Reference} \\
\colhead{($\mu$m)} & \colhead{(Jy)} & \colhead{} \\
}

\startdata
0.44 & 0.00012 & \citet{Myers1987} \\
0.55 & 0.00072 & \citet{Myers1987} \\
0.64 & 0.0034 & \citet{Myers1987} \\
0.79 & 0.014 & \citet{Myers1987} \\
1.24 & 0.0385 & 2MASS \\
1.25 & 0.12 & \citet{Myers1987} \\
1.65 & 0.43 & \citet{Myers1987} \\
1.66 & 0.15 & 2MASS \\
2.16 & 0.4 & 2MASS \\
2.20 & 0.7 & \citet{Myers1987} \\
3.40 & 1.29038 & \citet{Rebull2011} \\
3.45 & 1.6 & \citet{Myers1987} \\
3.60 & 2.1 & \citet{Cieza2009} \\
4.50 & 3.3 & \citet{Cieza2009} \\
4.60 & 8.68941 & \citet{Rebull2011} \\
4.80 & 3.3 & \citet{Myers1987} \\
5.80 & 9.6 & \citet{Cieza2009} \\
7.80 & 8.0 & \citet{Myers1987} \\
8.00 & 9.6 & \citet{Cieza2009} \\
8.70 & 7.0 & \citet{Myers1987} \\
9.50 & 7.00 & \citet{Myers1987} \\
10.10 & 9.0 & \citet{Myers1987} \\
10.30 & 10.0 & \citet{Myers1987} \\
11.60 & 12.0 & \citet{Myers1987} \\
12.00 & 16.6 & IRAS \\
12.00 & 24.02720 & \citet{Rebull2011} \\
12.50 & 16.0 & \citet{Myers1987} \\
20.00 & 26.0 & \citet{Myers1987} \\
22.00 & 37.5286 & \citet{Rebull2011} \\
25.00 & 37.6 & IRAS \\
60.00 & 58.4 & IRAS \\
100.00 & 47.0 & IRAS \\
350.00 & 1.68 & \citet{Andrews2005} \\
350.00 & 3.42 & \citet{Dent1998} \\
443.00 & 2.37 & \citet{Chandler1998} \\
443.00 & 1.81 & \citet{Andrews2005} \\
450.00 & 1.73 & \citet{Dent1998} \\
790.00 & 0.571 & \citet{Chandler1998} \\
800.00 & 0.353 & \citet{Dent1998} \\
863.00 & 0.28 & \citet{Andrews2005} \\
1100.00 & 0.138 & \citet{Dent1998} \\
1104.00 & 0.18 & \citet{Chandler1998} \\
1260.00 & 0.099 & \citet{Chandler1998} \\
1300.00 & 0.20 & \citet{Mott2001} \\
1927.00 & $<$0.16 & \citet{Chandler1998} \\
3400.00 & 0.0113 & \citet{Hogerheijde1997}
\enddata
\label{UnresPhot}

\end{deluxetable}

In addition to our CARMA data and archival HST imaging data we collected photometry for GV Tau from the literature to create a spectral energy distribution (SED). Because the components of the GV Tau binary are only separated by 1.2" the photometry from the literature for GV Tau is largely unresolved. We did however find a number of studies that spatially resolved the binary and provided photometry for each component. We also used VLT resolved near-infrared spectroscopy of the silicate feature for both components from \citet{Roccatagliata2011}. We list the resolved and unresolved photometry in Tables \ref{UnresPhot} and \ref{ResPhot} and plot the spectra in Figure \ref{specplot}. For our model fitting we ignored the uncertainties quoted in the literature and used a uniform 10\% uncertainty for each data point, although this value is somewhat arbitrary.

\begin{deluxetable}{cccc}
\tablecaption{Resolved photometry of GV Tau}
\tablewidth{0pt}
\tabletypesize{\tiny}
\tablehead{
\colhead{$\lambda$} & \colhead{F$_{\nu,S}$} & \colhead{F$_{\nu,N}$} & \colhead{Reference} \\
\colhead{($\mu$m)} & \colhead{(Jy)} & \colhead{(Jy)} & \colhead{}
}

\startdata
0.59 & 0.000437 & 0.000001 & \citet{Roccatagliata2011} \\
0.79 & 0.002365 & 0.000009 & \citet{Roccatagliata2011} \\
1.65 & 0.037 & 0.0021 & \citet{Roccatagliata2011} \\
1.65 & 0.402 & 0.028 & \citet{Leinert1989} \\
2.16 & 0.1329 & 0.0334 & \citet{Roccatagliata2011} \\
2.20 & 0.615 & 0.08 & \citet{Leinert1989} \\
3.45 & 0.838 & 0.631 & \citet{Leinert1989} \\
4.80 & 1.468 & 1.837 & \citet{Leinert1989} \\
1300 & 0.0404 & 0.0443 & This work \\
1300 & 0.0467 & 0.0438 & \citet{Guilloteau2011} \\
2700 & 0.0091 & 0.0105 & \citet{Guilloteau2011} \\
36000 & 0.0011 & 0.0001 & \citet{Reipurth2004}
\enddata
\label{ResPhot}

\end{deluxetable}

\citet{Reipurth2004} used the Very Large Array (VLA) A configuration to observe GV Tau at 3.6 cm with 0.3" resolution and resolved the components of the binary. The 3.6 cm emission detected towards the southern component appears to trace an outflow and is likely not thermal dust emission, while the emission detected towards the northern component is consistent with thermal dust emission with a spectral index of 2. Given our current data we cannot be certain of the origin of the 3.6 cm emission from either source so we exclude the point from our modeling for the time being. We intend to follow up on this feature in a future paper.

The left panel of figure \ref{specplot} shows a plot of the photometry for GV Tau in which the binary was not resolved, along with the sum of the photometry for each component. The composite GV Tau N and S photometry matches the unresolved data well at wavelengths longer than 3 $\mu$m. At shorter wavelengths the composite photometry falls below the unresolved photometry. This is likely because the unresolved data use a larger aperture and thus includes more of the nebulosity which is present in near-infrared images of GV Tau. In our modeling, described below, we fit individual protostar models to the resolved photometry for each component. We also include the unresolved photometry from 12 - 100 $\mu$m as upper limits to constrain the modeling as we do not have resolved photometry in that range.

\section{Modeling}

We follow the same modeling procedure as \citet{Eisner2005} and \citet{Eisner2012} using a grid of models described below.

\subsection{Input Density Distributions}

Our models include a central protostar surrounded by a circumstellar disk and an envelope with an outflow cavity. We provide below further details of the structure of each component and the parameters that were varied to create our grid of models.

\subsubsection{Protostar}

We use a central protostar with a temperature of 4000 K and a mass of 0.5 M$_{\odot}$ for both protostars in the GV Tau system. This is consistent with previous studies of GV Tau, which find a mass and temperature of 0.5 M$_{\odot}$ and 3800 K for GV Tau S and of 0.8 M$_{\odot}$ and 4100 K for GV Tau N \citep{Doppmann2008}. We allow the luminosity of the protostar to be 1, 3 or 6 L$_{\odot}$, and calculate the radius of the protostar accordingly, assuming that the protostar is a spherical blackbody. Our selected luminosities are compatible with previous luminosity measurements by \citet{White2004}, \citet{Doppmann2005}, and \citet{Prato2009}. While \citet{Doppmann2008} measured lower luminosities for the protostars (0.3 L$_{\odot}$ and 0.6 L$_{\odot}$), the assumed age in that study may be too old, thus pushing the luminosity down. We discuss this further in Section \ref{GVTau_age}. The spectrum of the protostar is also assumed to be that of a spherical blackbody with a temperature of 4000 K.

\subsubsection{Envelope}

We model the density distribution of the protostellar envelope using the solution for a rotating collapsing envelope \citep{Ulrich1976},
\begin{equation}
\rho_{env}(r,\mu) = \frac{\dot{M}}{4\pi}\left(G M_* r^3\right)^{-\frac{1}{2}} \left(1+\frac{\mu}{\mu_0} \right)^{-\frac{1}{2}} \left(\frac{\mu}{\mu_0}+2\mu_0^2\frac{R_c}{r}\right)^{-1},
\end{equation}
where $r$ and $\theta$ are defined in the typical sense for spherical coordinates centered on the protostar, and $\mu = \cos{\theta}$. The parameter $\dot{M}$ is the accretion rate of the envelope onto the protostar, and can be calculated from the total envelope mass by integration over all space. $R_c$ is the centrifugal radius of the envelope, interior to which the density distribution begins to significantly flatten due to rotation. $\mu_0 = \cos{\theta_0}$ is the initial angle of the infalling material and can be solved numerically from the equation \citep{Ulrich1976},
\begin{equation}
\frac{r}{R_c} = \frac{1-\mu_0^2}{1-\mu / \mu_0}.
\end{equation}
Finally, we truncate the envelope at a an outer radius, $R_{\rm env}$, and at a fixed inner radius, $R_{\rm in}$.

The inner radius of the envelope in our models is fixed at a distance of 0.1 AU, consistent with previously measured inner disk radii for our range of model luminosities \citep[e.g.][]{Eisner2007}, while the total envelope mass, $M_{\rm env}$, and the outer radius of the envelope, $R_{\rm env}$, are left as free parameters to be varied in our grid. We allow $M_{\rm env}$ to take values of $1 \times 10^{-6}$, $5 \times 10^{-6}$, $1 \times 10^{-5}$, $5 \times 10^{-5}$, $1 \times 10^{-4}$, and $5 \times 10^{-4}$ M$_{\odot}$, and $R_{\rm env}$ is selected from 60, 90, 300 and 1000 AU. Although the centrifugal radius, $R_c$, is a free parameter, we fix it to be equal to the radius of the protoplanetary disk, described below. It can take values of 30, 60, 100, or 300 AU. We allow values of $R_c$ larger than the projected separation of the protostars (170 AU) because of the possibility that the actual separation is much larger.

We also give the envelope an outflow cavity, the location of which is determined by
\begin{equation}
z > 1 \mbox{AU} + r^{\zeta}
\end{equation}
Inside the outflow cavity, the density of the envelope is reduced by a scale factor, $f_{\rm cav}$. We leave $f_{\rm cav}$ as a free parameter which is allowed to take values of 0.05, 0.2, and 1, and hold $\zeta$ fixed at a value of 1.0. While it would be nice to vary $\zeta$, computational limitations dictate that we hold some parameters fixed. The parameter study in \citet{Eisner2012} suggests that $\zeta$ primarily affects the overall flux scaling of the 1.3 mm visibilities as well as the offset between the scattered light emission and the protostar. This would suggest that $\zeta$ is degenerate with the disk mass, which is largely responsible for the overall flux scaling of the 1.3 mm visibilities, however $\zeta$ only affects this scaling on the order of 10\% so we do not believe that it produces a significant error in our disk mass measurements. $\zeta$ may, however have a significant effect on inclination and position angle, but the astrometry errors of our scattered light image likely overshadow this error.

\subsubsection{Protoplanetary Disk}

To model the protoplanetary disk, we use the standard prescription for a flared viscous accretion disk,
\begin{equation}
\rho_{disk}(r,z) = \rho_0 \left(\frac{r}{1 AU}\right)^{-\alpha} \exp\left(-\frac{1}{2}\left[\frac{z}{h(r)}\right]^2\right),
\end{equation}
\begin{equation}
h(r) = h_0\left(\frac{r}{1 AU}\right)^{\beta},
\end{equation}
with $r$ and $z$ defined in the usual sense for cylindrical coordinates. $\rho_0$ is the density of the disk at the midplane at a radius of 1 AU, and can be calculated from the total disk mass, $M_{\rm disk}$, by integrating the disk density over all space. $h_0$ is the scale height of the disk at 1 AU. We truncate the disk at a given outer radius, $R_{\rm disk}$, and inner radius, $R_{\rm in}$. 

In our models we fix $\beta$ at a value of 58/45 (or 1.29) as found by \citet{Chiang1997} for a flared accretion disk in hydrostatic equilibrium. Viscous accretion theory specifies that $\alpha = 3(\beta - \frac{1}{2}) = $ 71/30 (or 2.37) \citep{Shakura1973}. For these values of $\alpha$ and $\beta$ the surface density is proportional to $r^{-1.08}$. We take the scale height at 1 AU, $h_0$, to be 0.15 AU, and hold the inner radius of the disk fixed at 0.1 AU, consistent with measurements of the inner disk radius of T Tauri stars for the range of luminosities we chose \citep[e.g.][]{Eisner2007}. The total disk mass, $M_{\rm disk}$, and the outer radius of the disk, $R_{\rm disk}$, are left as free parameters. In our grid we allow $M_{\rm disk}$ to take values of $1 \times 10^{-6}$, $5 \times 10^{-6}$, $1 \times 10^{-5}$, $5 \times 10^{-5}$, and $1 \times 10^{-4}$ M$_{\odot}$, and we let $R_{\rm disk}$ vary between 30, 60, 100 and 300 AU. We again allow large disk radii because the actual separation of the protostars may be larger than the projected separation. We do not allow models in which $R_{\rm disk} > R_{\rm env}$.

\subsubsection{Summary of Model Parameters}

The model we employ includes a significant number of free parameters, and creating a grid of models that can fully explore the parameter space of these models is not practical. This is especially true because of the significant amount of computational time required to generate a single model, meaning that our grid must be relatively coarse out of necessity. Instead we focus on the subset of the free parameters which are particularly important for determing the best model fit to the data. We hold $h_0$ and $\zeta$ constant so that we can explore more values for other parameters. The parameter study from \citet{Eisner2012} suggests that these parameters have a smaller influence on models compared with other parameters. The free parameters in our models are $M_{\rm disk}$, $R_{\rm disk} = R_c$, $M_{\rm env}$, $R_{\rm env}$, $L_{\rm star}$, and $f_{\rm cav}$.

\subsection{Opacity}

We calculate the opacity of dust grains in our model following the prescription of \citet{Pollack1994}. Our dust grains are composed by volume of a mixture of 38$\%$ astronomical silicates, $3\%$ troilite, $29\%$ organics, and $30\%$ water ice. Optical constants for the astronomical silicates, troilite, organics, and water ice were taken from \citet{Draine2003}, \citet{Begemann1994}, \citet{Pollack1994}, and \citet{Hudgins1993} respectively. We reduce the amount of water ice relative to the other constituents when compared with the \citet{Pollack1994} recipe to account for the high temperatures which would vaporize much of the water ice in the inner regions of the disk. We calculate the optical properties of the mixed grains using the Bruggeman mixing rule. We calculate the absorption and scattering opacities from the optical properties of the combined grains assuming that the grains are spherical and using the code BHMIE \citep{Bohren1983}.

We assume that the dust in our models follows a power law grain size distribution, $n(a) \propto a^{-p}$, between some minimum and maximum grain size. \citet{Mathis1977} find that $p = 3.5$ for the interstellar medium, and several investigators have found that the collisional cascade in debris disks also results in $p \approx 3.5$ \citep[e.g.][]{Dohnanyi1969}. The power law exponent in Class I disks is not well known, so we assume $p = 3.5$. We assume a minimum grain size of 0.005 $\mu$m for all of our opacities. The dust in the envelope of our models always uses opacities with a maximum grain size of 1 $\mu$m, roughly consistent with dust grains in the interstellar medium. We, however, allow the maximum dust grain size, $a_{max}$, in the disk to take values of 1 $\mu$m, 10 $\mu$m, and 1 mm so that we might explore grain growth in the disk of our targets.

\begin{figure*}[h!]
\centering
\includegraphics[width=6in]{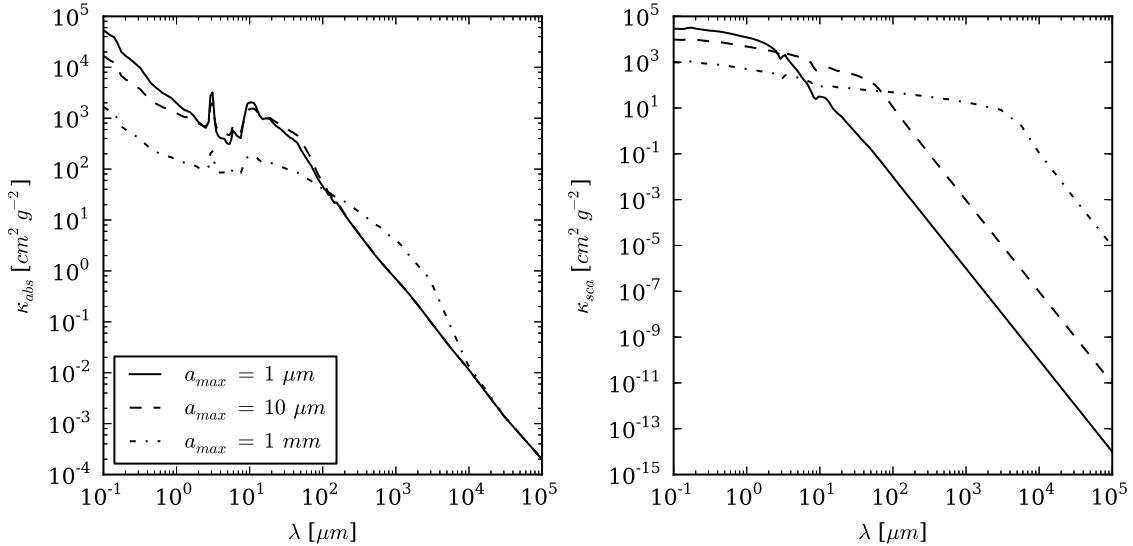}
\caption{Opacities we use in our modeling, with three different maximum grain sizes. We plot the absorption and scattering coefficients for the opacities in the left and right panels respectively. The behavior of our opacities with increasing $a_{max}$ agrees qualitatively with both the opacities used by \citet{DAlessio2001}.}
\label{opacplot}
\end{figure*}

Our opacities for dust grains with a maximum grain size of 1 $\mu$m are in good agreement with the \citet{Ossenkopf1994} dense ($n = 10^5$) protostellar core opacities, differing by at most a factor of two. This is well within the degree to which we know the composition of interstellar dust grains. We also compare our opacities qualitatively with the opacities of \citet{DAlessio2001}, who follow a similar recipe, and find that our results are in good qualitative agreement for different grain size distributions. We plot the effect of changing the maximum dust grain size in Figure \ref{opacplot}. Changing the relative abundances of the constituent grains produced smaller effects on resulting opacities.

\subsection{Radiative Transfer Codes}

We use a combination of the three dimensional Monte Carlo dust radiative transfer codes Hyperion \citep{Robitaille2011} and RADMC-3D to create our grid of spectra, images and visibilities of model protostars. For a given dust density distribution, we use Hyperion to perform a Monte Carlo simulation to calculate the dust temperature throughout the grid. We then use RADMC-3D raytracing to create synthetic spectra and images of the model from the density and temperature distributions. Finally, we Fourier transform synthetic 1.3mm images to create visibilities. We describe the basic functioning of these codes, as well as our rationale for using two in tandem, below.

Hyperion runs Monte Carlo simulations to determine the temperature structure of a given protostellar model using the iteration method proposed by \citet{Lucy1999}. Photons are propagated through the density grid, being absorbed and re-emitted as they go, and after all of the photons have escaped the grid, the temperature is computed, and the simulation is carried out again until the temperature converges to a satisfactory level. For our simulations, each iteration used $10^5$ photons, which we found to give a good balance between accuracy in the temperature measurement and time required to run a single model simulation. Each simulation usually requires $\sim$10 iterations to converge, for a total of about $10^6$ photons for each thermal simulation.

Our models tend to include very high density regions (e.g. the disk midplane) into which few photons travel. To improve the signal-to-noise of the temperature measurement we allow Hyperion to use the Partial Diffusion Approximation (PDA) to more accurately calculate the temperature in these high optical depth regions following each iteration. Furthermore, if a photon does wander into these regions of the grid it can end up being trapped in the high density cell, which significantly slows down the calculation. To circumvent this problem we employ the Modified Random Walk \citep[MRW;][]{Min2009,Robitaille2010} method which can speed up these trapped photons by allowing the photons to diffuse out of the cell in a single step rather than hundreds or thousands.

Images and SEDs are computed for our models using raytracing of the dust thermal emission. To account for scattered light emission, which can contribute a significant fraction of the signal at short wavelengths, we run scattered light simulations in which monochromatic photons are propagated through the grid and allowed to scatter until they are absorbed. The scattering phase function can then be determined by the scattering properties of the photons and included in the raytracing algorithm to quickly create images and SEDs from the models. The scattered light simulations are run with $10^4$ photons at each wavelength in an SED generated and $10^5$ photons for each image, which was found to give good signal-to-noise in our models. We calculate the images and SEDs for inclinations from $0^{\circ}$ to $90^{\circ}$ at intervals of $5^{\circ}$, and we vary the position angle from $0^{\circ}$ to $360^{\circ}$ in intervals of $10^{\circ}$.

We elected to use a combination of both codes because we frequently found that the run time for each thermal simulation was dominated by photons being trapped in high optical depth regions, so the MRW and PDA procedures significantly sped up the simulations. At the time when we created our grid of models, only Hyperion employed both of these procedures and thus we elected to use Hyperion for the thermal simulation portion of our modeling. Conversely, raytracing for both thermal emission and scattered light is the fastest method for producing spectra and images, and at the time when we created our grid of models, only RADMC-3D offered raytracing for scattered light. We note, however, that since we ran our model grid RADMC-3D has been updated to include the MRW algorithm and has a PDA module under development. Furthermore, Hyperion may include raytracing for scattered light in the future \citep{Robitaille2011}.

\subsection{Model fitting}

We fit our models to all three datasets (SED, HST scattered light image, and CARMA visibilities) simultaneously with a weighted least squares fit. For each individual dataset we calculate $\chi^2$ for the corresponding component of each model. We then combine the separate $\chi^2$ measurements into one weighted least squares parameter:
\begin{equation}
X^2 = \left(w_{spec} \frac{\chi^2_{spec}}{min(\chi^2_{spec})} + w_{im} \frac{\chi^2_{im}}{min(\chi^2_{im})} + w_{vis} \frac{\chi^2_{vis}}{min(\chi^2_{vis})}\right) / (w_{spec} + w_{vis} + w_{im}).
\end{equation}
It is important to note that our goodness of fit parameter $X^2$ is not a true $\chi^2$ statistic and cannot be used in a statistically rigorous way.

We use our $X^2$ metric rather than a true $\chi^2$ to determine the best fit to the data so that we have the ability to change the weight given to each dataset. If we were to use a true $\chi^2$ metric, our fits would be dominated by the imaging data, which has a significantly larger number of data points than the visibilities or SED. The primary goal of this study is to constrain disk properties from our datasets. Fitting with a true $\chi^2$, however, would place most of the weight on the scattered light images, and scattered light images do not trace the disk properties as well as the visibilities or SED. Millimeter wavelength measurements are particularly sensitive to the dust mass of a system, so fits with larger weight given to the 1.3mm visibilities may more faithfully reproduce the distribution of denser material. Our $X^2$ statistic allows us to put more weight on the visibilities and make our fitting more sensitive to disk mass.

Our $X^2$ metric allows us to explore how consistent our best fit parameters are with each dataset individually. If a parameter remains relatively constant as we give each dataset significantly more weight than the others, that would suggest that the parameter is consistent with each of the datasets and is well constrained. Conversely, those parameters that vary significantly with different weights are likely not well constrained and may suggest that a more complex model is needed to fully explain the complete dataset.

We fit our models to the data for each component of the binary separately. In order to do this we fit the model spectrum to only the resolved spectrum of each source, while the unresolved data are used as upper limits. We also split the HST image into smaller sub-images which only contain the appropriate source, and split the visibilities as described above to obtain the visibilities for an individual source. However we do verify that our best-fit models also provide good fits to the composite imaging, visibility and photometric data.

\section{Results}

\begin{deluxetable}{cccccccccccccccc}
\centering
\tablecaption{Best fit models.}
\tablewidth{0pt}
\tabletypesize{\tiny}
\tablehead{
\colhead{} & \colhead{Source} & \colhead{$w_{mm}$} & \colhead{$w_{NIR}$} & \colhead{$w_{SED}$} & \colhead{$X^2$\tablenotemark{a}} & \colhead{$M_{disk}$} & \colhead{$R_{disk}$} & \colhead{$M_{env}$} & \colhead{$R_{out}$} & \colhead{$a_{max}$\tablenotemark{b}} & \colhead{$f_{cav}$} & \colhead{$L_*$} & \colhead{$i$} & \colhead{$PA$} \\
\colhead{} & \colhead{} & \colhead{} & \colhead{} & \colhead{} & \colhead{} & \colhead{(M$_{\odot}$)} & \colhead{(AU)} &\colhead{(M$_{\odot}$)} & \colhead{(AU)} & \colhead{($\mu$m)} & \colhead{} & \colhead{(L$_{\odot}$)} & \colhead{$(^{\circ})$} & \colhead{$(^{\circ})$}
}
\startdata
a & GV Tau N & 10 & 1 & 1 & 1.2 & $5\times10^{-5}$ & 30 & $5\times10^{-5}$ & 300 & 1 & 0.2 & 3 & 30 & 200 \\
b & GV Tau N & 1 & 1 & 10 & 1.2 & $1\times10^{-4}$ & 30 & $1\times10^{-5}$ & 90 & 1000 & 0.2 & 3 & 35 & 200 \\
c & GV Tau N & 1 & 1 & 1 & 1.5 & $1\times10^{-4}$ & 30 & $5\times10^{-5}$ & 300 & 1 & 0.2 & 3 & 30 & 200 \\
d & GV Tau N & 1 & 10 & 1 & 1.7 & $1\times10^{-4}$ & 30 & $5\times10^{-5}$ & 300 & 10 & 0.2 & 3 & 25 & 200 \\
\hline
a & GV Tau S & 10 & 1 & 1 & 1.1 & $5\times10^{-5}$ & 30 & $1\times10^{-5}$ & 300 & 10 & 1 & 6 & 55 & 160 \\
b & GV Tau S & 1 & 1 & 10 & 1.1 & $1\times10^{-4}$ & 30 & $1\times10^{-5}$ & 300 & 1000 & 1 & 6 & 55 & 160 \\
c & GV Tau S & 10 & 1 & 10 & 1.3 & $1\times10^{-4}$ & 30 & $1\times10^{-5}$ & 300 & 1000 & 1 & 6 & 55 & 160 \\
d & GV Tau S & 1 & 1 & 1 & 1.4 & $1\times10^{-4}$ & 30 & $1\times10^{-5}$ & 300 & 1000 & 1 & 6 & 55 & 160 \\
\enddata
\tablenotetext{a}{Note that $X^2$ is a measure of the goodness of fit of a model, as defined by Equation 6, and not a true $\chi^2$.}
\tablenotetext{b}{The parameter $a_{max}$ is the maximum dust grain size for the opacity used in the disk. The maximum dust grain size in the envelope is held constant at 1 $\mu$m.}
\label{best_fits}
\end{deluxetable}

\begin{figure*}
\centering
\includegraphics[width=6in]{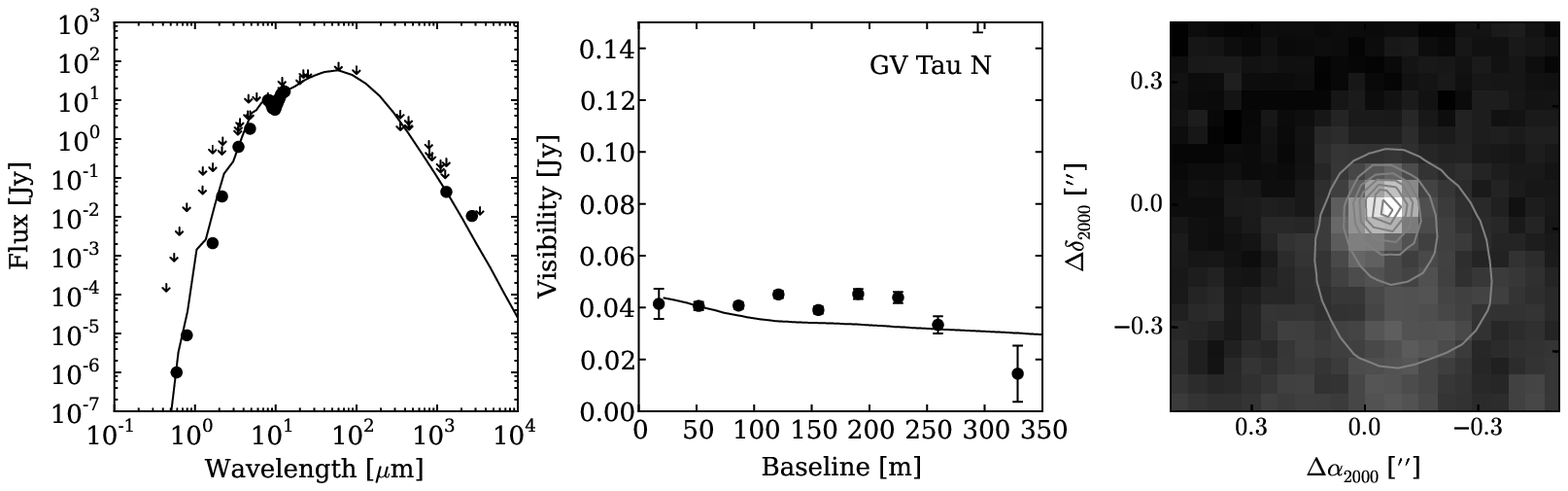}
\includegraphics[width=6in]{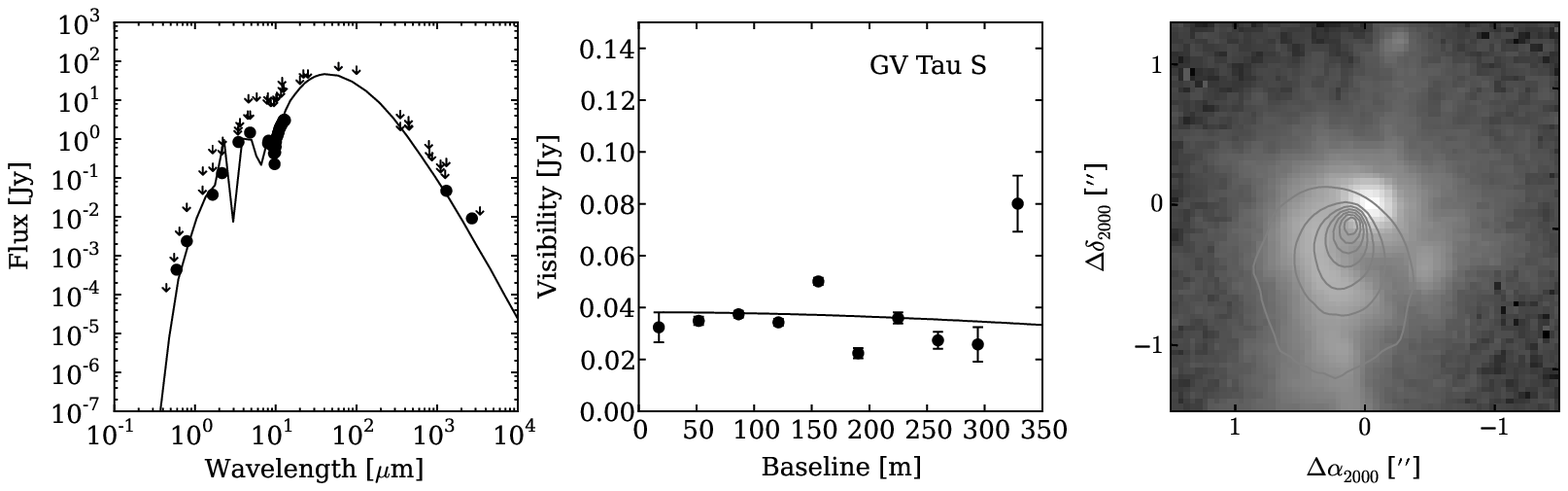}
\includegraphics[width=6in]{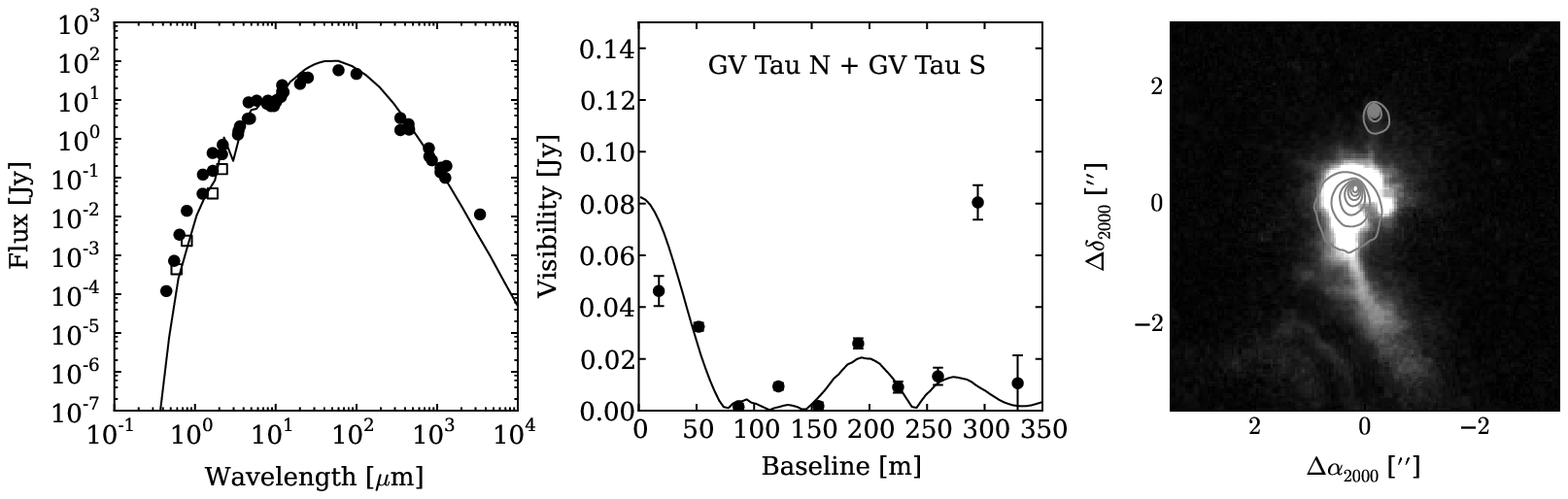}
\includegraphics[width=4in]{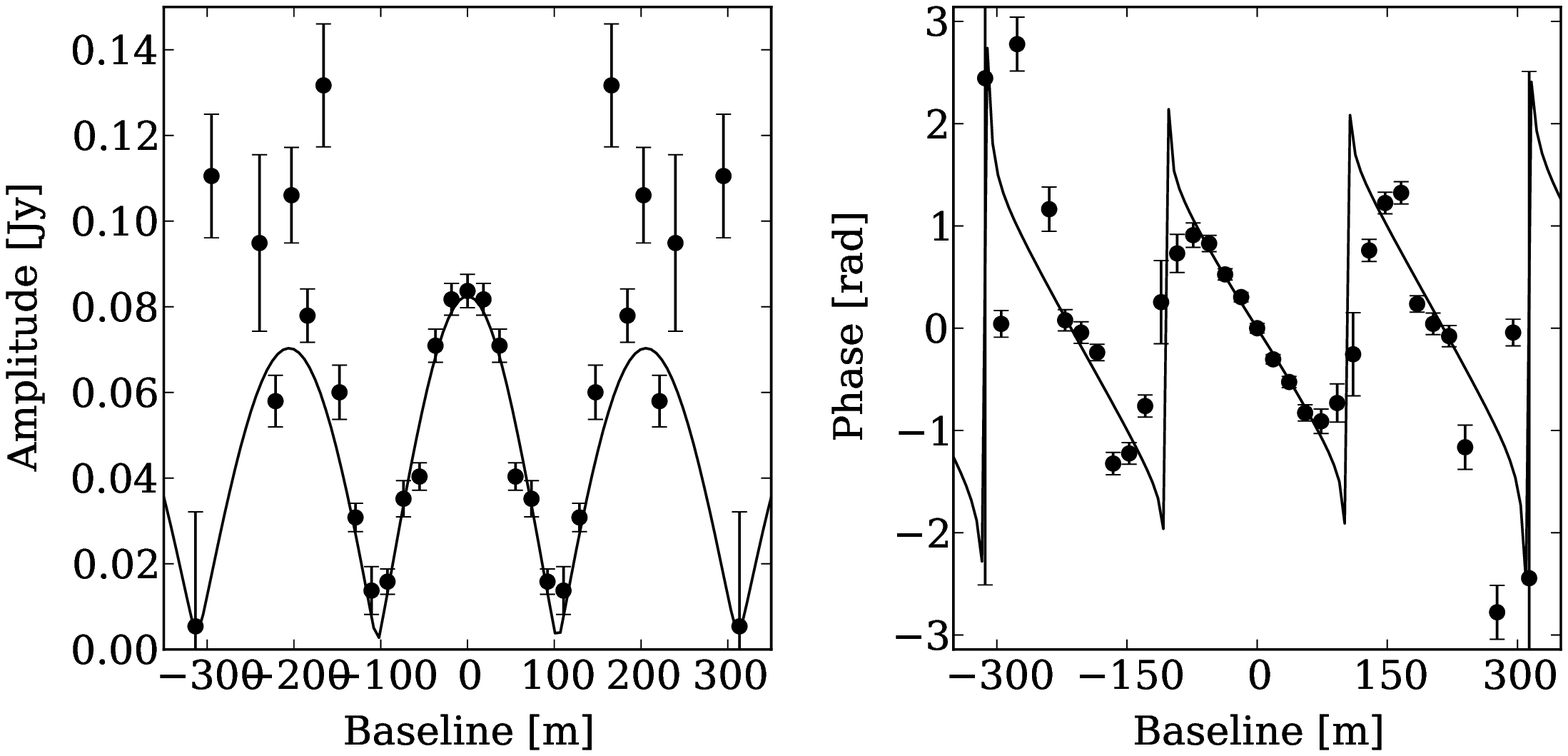}
\caption{Our data, shown as solid points, with the best fit models, shown as lines, for GV Tau N and S overplotted. The first three rows show broadband SEDs in the left panel, 1.3 mm visibilities in the middle panel, and 0.8 $\mu$m scattered light images in the right panel. Panels which show the resolved spectra of GV Tau N or S also show the unresolved spectrum of the system as upper limits. The first row shows the data and model for GV Tau N, while the second row shows the same for GV Tau S. The third and fourth rows show the combined GV Tau dataset with the sum of the best fit north and south models plotted on top. The fourth row shows the visibilities, both amplitude and phase, averaged along the binary axis.}
\label{best_fit_opt}
\end{figure*}

\begin{figure*}
\centering
\includegraphics[width=6in]{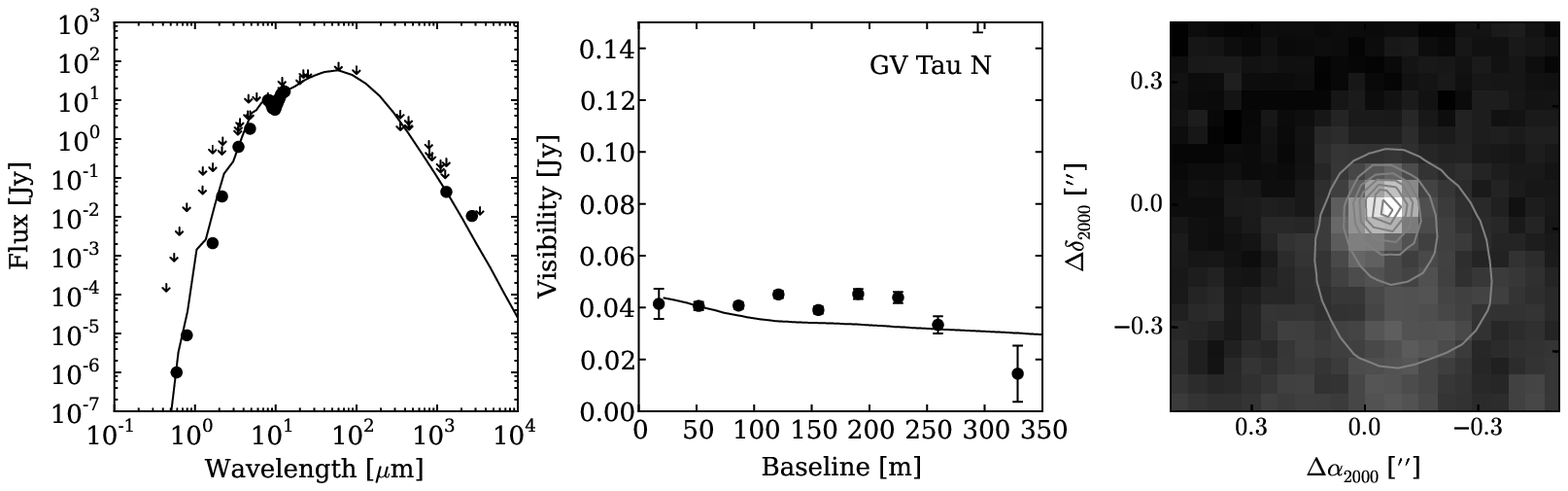}
\includegraphics[width=6in]{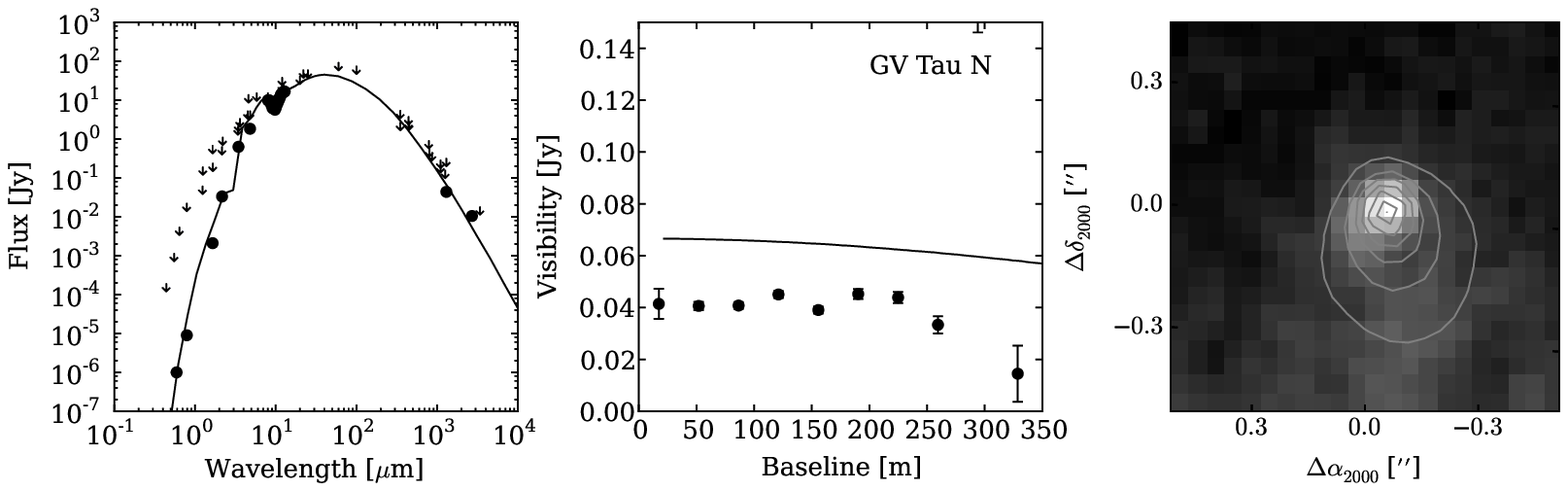}
\includegraphics[width=6in]{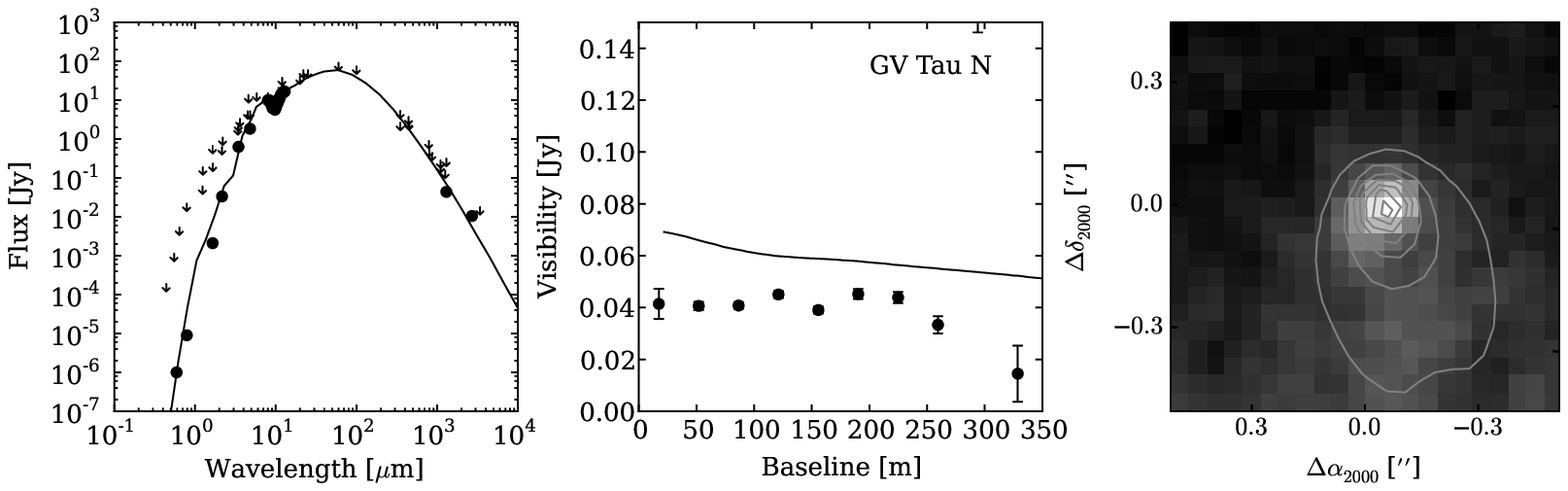}
\includegraphics[width=6in]{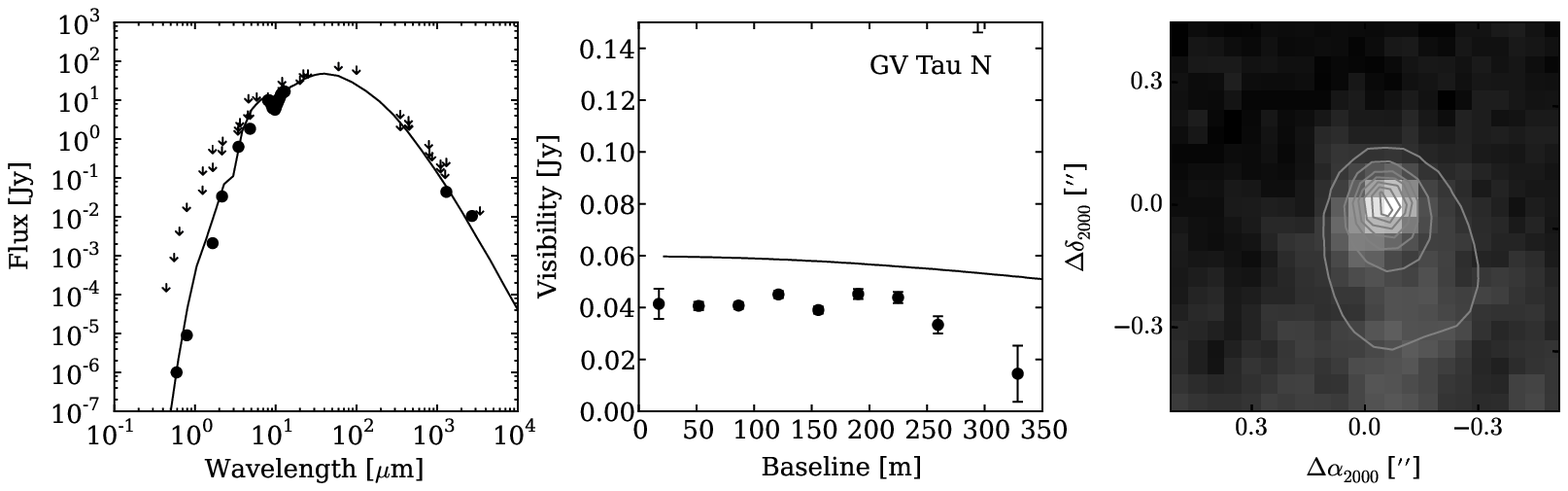}
\caption{Models for GV Tau N which show good fits to the data. See Figure \ref{best_fit_opt} for more information about each panel. Such models can be used to determine how robust our determination of each parameter is, as well as how uncertain our measurements may be. Each row shows the model for the corresponding row for GV Tau N in Table \ref{best_fits}. Like Table \ref{best_fits}, the plots are ordered by increasing $X^2$.}
\label{vary_N}
\end{figure*}

\begin{figure*}[h!]
\centering
\includegraphics[width=6in]{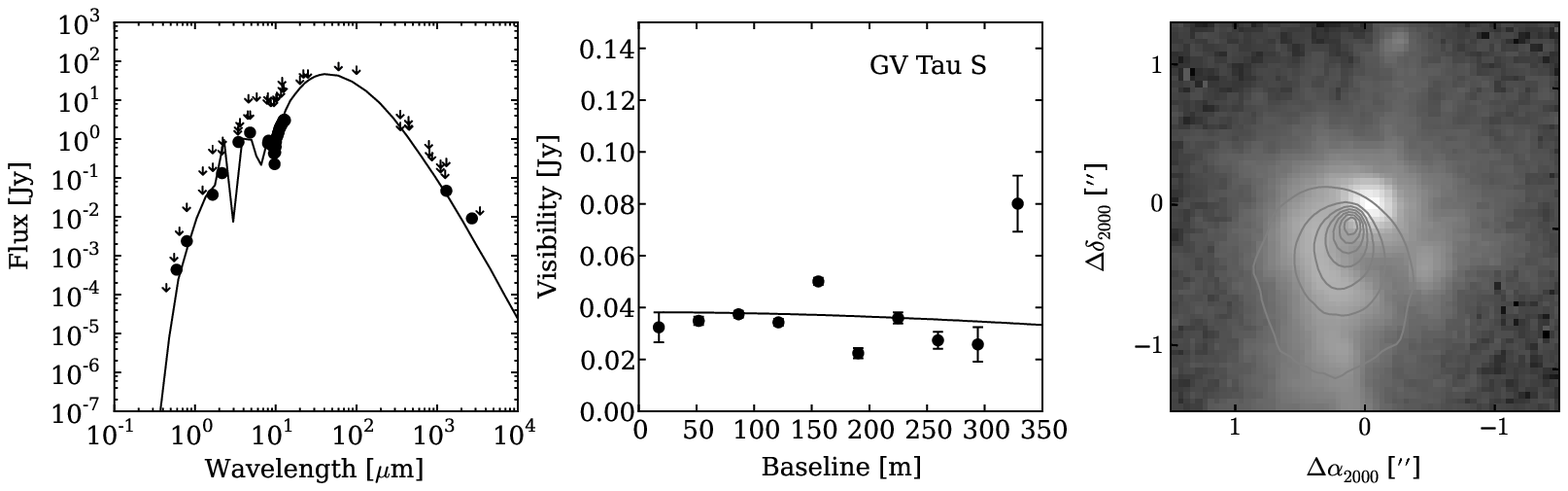}
\includegraphics[width=6in]{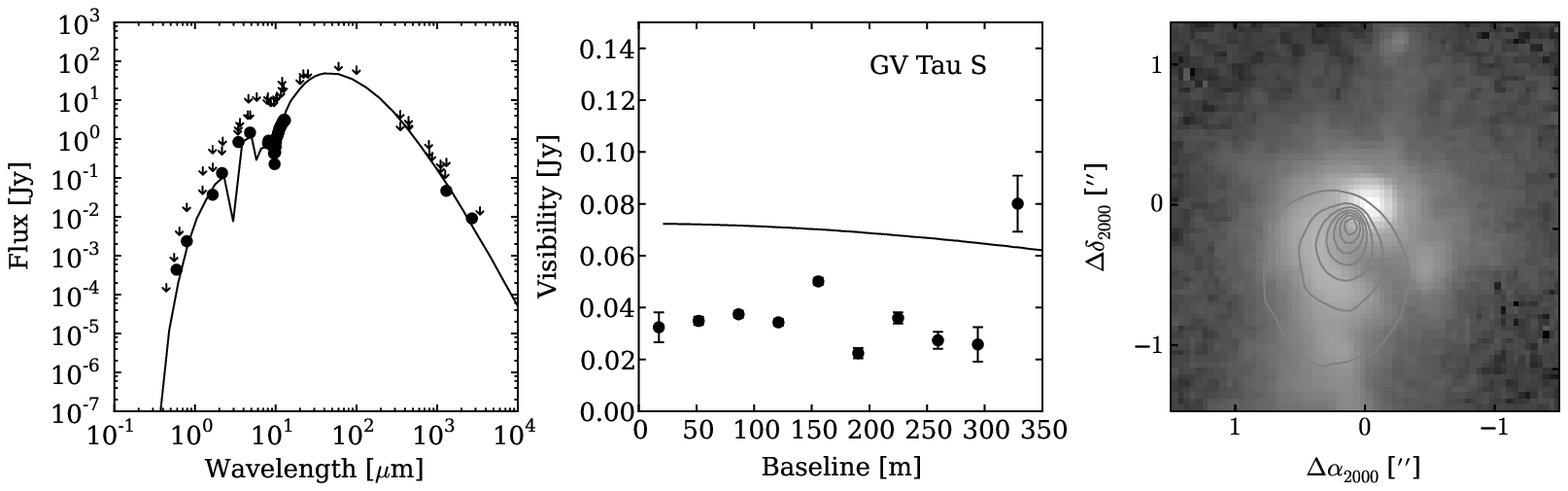}
\includegraphics[width=6in]{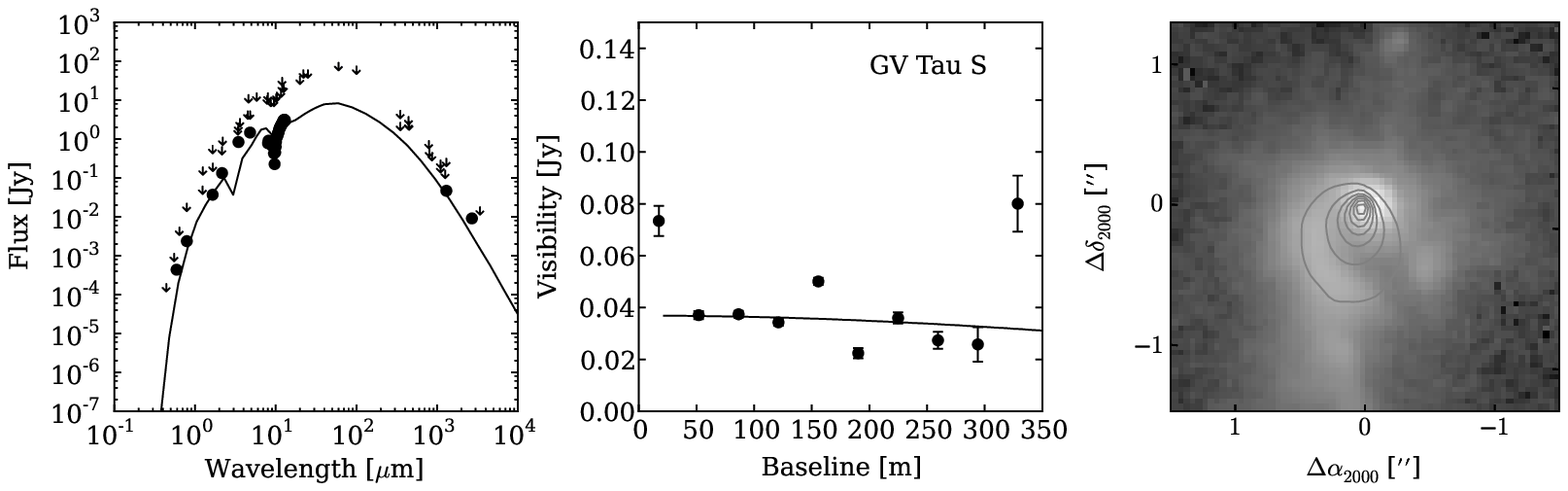}
\caption{Models which fit the data for GV Tau S well. See Figure \ref{best_fit_opt} for more information about each panel. The first row is our best fit model, as described by row ``a" for GV Tau S in Table \ref{best_fits}, and the second row shows the model from rows b-d, as the rows are identical. In the third row we show a good model fit which was tuned by hand to plausibly reproduce the 8-13 $\mu$m visibilities from \citet{Roccatagliata2011} while maintaining a good fit to our datasets (see Figure \ref{GVTauS_spec_int} for further details).}
\label{vary_S}
\end{figure*}

\begin{figure*}[h!]
\centering
\includegraphics[width=6in]{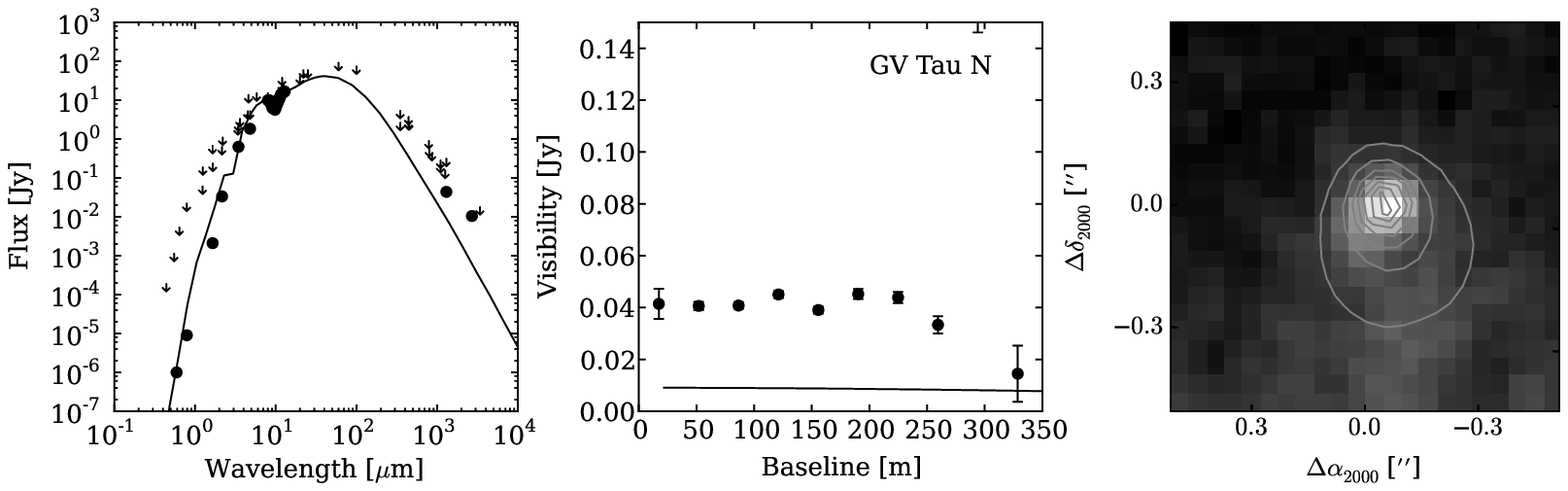}
\includegraphics[width=6in]{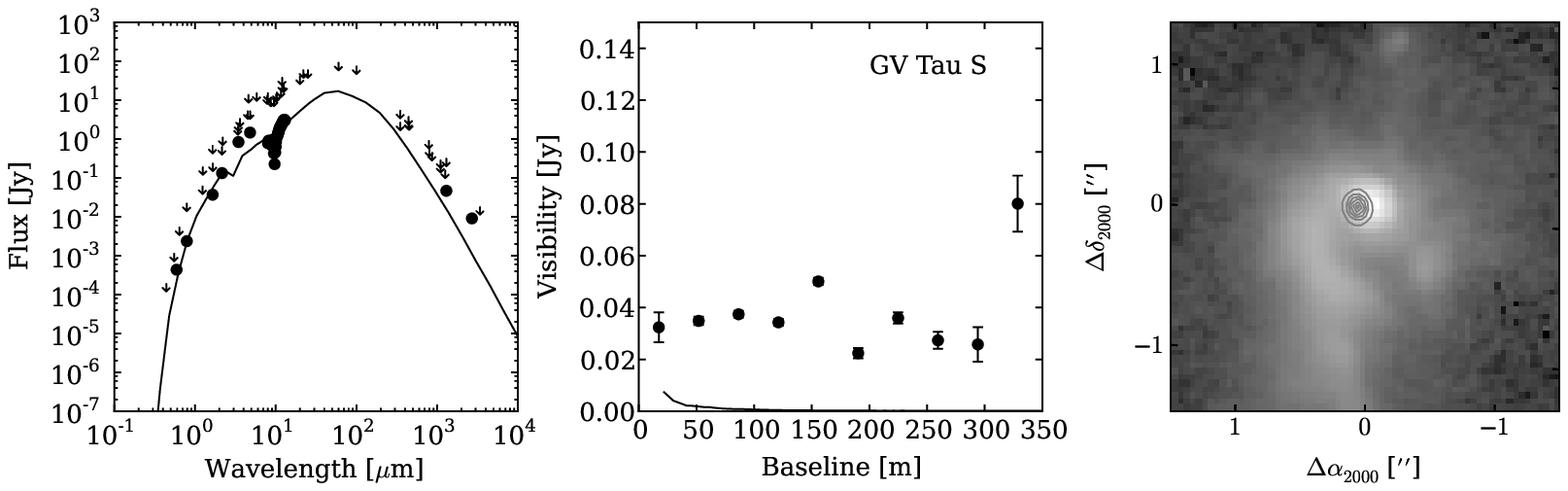}
\caption{Models for both sources which do not fit the data well. See Figure \ref{best_fit_opt} for more information about each panel. We show a model for GV Tau N with $X^2 = 2.0$ in the top row and a model for GV Tau S with $X^2 = 1.7$ in the bottom row. Models with $X^2$ above our thresholds, of 2 for GV Tau N and 1.7 for GV Tau S, no longer reproduce the data well, as demonstrated by the poor fits of these models.}
\label{bad_fits}
\end{figure*}

We list the best fit model parameters for GV Tau N and S in Table \ref{best_fits}, ordered by increasing $X^2$. We plot the synthetic data for the best fit models, as described in row ``a" of Table \ref{best_fits}, in Figure \ref{best_fit_opt}. We also plot the sum of the best fit models against the full binary dataset to show that we fit all of the composite data. We give the mm data more weight, as described above, in our best-fit model.  However we explore other weighting schemes to determine whether best-fit parameters are consistent with each dataset individually.  If parameters change as weights are varied, that implies that some model parameters may produce degenerate effects, or that a more complex model may be required to fit the combined dataset well. We list all models with $X^2 < 2$ for GV Tau N and $X^2 < 1.7$ for GV Tau S in Table \ref{best_fits}, and those models are plotted in Figures \ref{vary_N} \& \ref{vary_S} in order of increasing $X^2$. In Figure \ref{bad_fits} we show the models with the next-lowest $X^2$ values. These models are clearly unsuitable fits to the data, so we do not consider them, or models with still higher $X^2$ values, further.

We find that the disk dust masses of the best fit models for GV Tau N and S are each $5.0 \times 10^{-5}$ M$_{\odot}$. By examining the fit quality across our model grid, we provide qualitative estimates of the acceptable range of parameter values. A disk mass a factor of two larger than our best fit model can still reproduce our dataset well (Figures \ref{vary_N} \& \ref{vary_S}), while a disk mass five times lower cannot (Figure \ref{bad_fits}). While not statistically rigorous, we therefore estimate that the disk masses for GV Tau N and S are constrained to within a factor of 2.

Opacity is likely a large source of additional uncertainty on our disk mass measurement because millimeter flux measurements are sensitive to the product of mass and opacity. We allowed the maximum dust grain size to vary in an attempt to constrain the opacity, but we are unable to definitively determine the properties of the opacity. Both protostars can be fit by models with maximum grain sizes which range across the spectrum of allowed values. Interestingly, regardless of which opacity law is used, the disk mass is measured to be the same. One might expect that for the larger values of $a_{max}$ that we consider we would measure a lower $M_{disk}$ because the 1.3mm opacity is higher. For larger values of the 1.3mm opacity however, it turns out that our best fit model disks are significantly more optically thick than for lower 1.3mm opacities. This means that more mass is needed than is otherwise expected to reproduce the 1.3mm flux. It is likely because of this high optical depth that our modeling has difficulties in constraining the opacity in the disk. Furthermore, for $a_{max} >> 1$ mm the dust opacity at 1.3 mm will drop, also allowing for a larger inferred disk mass. Changing the opacity parameters also has different effects at different wavelengths, so models may require the mass to remain constant to fit the combined dataset, even as the opacity is varied.

We are, however, able to place a constraint on the radii of the GV Tau N and S disks. Our modeling shows that both protostars strongly favor models with $R_{disk}=30$ AU. This is the smallest radius allowed in our model grid, so it is possible the the true disk radii are, in fact, smaller than our best fit models suggest. It is also possible that the disk radius could be somewhat larger than 30 AU, as the next smallest radius in our grid is $R_{disk}=60$ AU. A disk with a radius of $\sim60$ AU would have an extent of $\sim0.8"$ and would be marginally resolved by our 1.3 mm visibilities. Our data, however, suggest that the disks are unresolved, so we conlcude that $R_{disk} < 30$ AU.

The parameters of the envelopes also appear to be somewhat constrained by our modeling, however we also find some degeneracy between envelope mass and radius. The best fit envelope dust mass is $5.0 \times 10^{-5}$ M$_{\odot}$ for GV Tau N and $1.0 \times 10^{-5}$ M$_{\odot}$ for GV Tau S, and both protostars have $R_{env} = 300$ AU. Furthermore, our modeling suggests that $f_{cav} = 0.2$ for GV Tau N and $f_{cav} = 1$ for GV Tau S. It is, however, possible to decrease (or increase) both M$_{env}$ and R$_{env}$ while maintaining the quality of fit to the data. This is unsurprising because our millimeter visibilities have very limited sensitivity to faint extended structures, particularly those close to or larger than 1000 AU. As such, we suggest that these envelope parameters should be treated with caution.

Furthermore, from our modeling we are able to marginally constrain the viewing geometry of the GV Tau system. We find that the best fit models for GV Tau N suggest an inclination of $30^{\circ}$ while they suggest an inclination of $55^{\circ}$ for GV Tau S. Similarly, we find a position angle of $200^{\circ}$ for GV Tau N and $160^{\circ}$ for GV Tau S. These parameters however, are dependent on the astrometry of the scattered light image, which we find to be quite uncertain. If we adjust the astrometry within the bounds allowed by our uncertainty, we find that the position angle of neither source is well constrained by our modeling. If we consider the uncertainty in the astrometry, as well as the variations of best fit inclinations as we change the weighting of our datasets, we estimate that we could vary our best fit inclinations by up to $20^{\circ}$ and still find acceptable fits to the dataset. The inclination is better constrained than the position angle because the SED can provide an additional constraint only on the inclination, while the position angle is constrained almost entirely by the scattered light imaging.

Finally, we find that the luminosity of each protostar is constrained by our modeling, with an accuracy limited by the sparse sampling of the parameters in our grid. GV Tau N very strongly prefers a luminosity of 3 L$_{\odot}$ while GV Tau S tends to favor a luminosity of 6 L$_{\odot}$. We have also found a model with a $L_{star} = 1.5$ L$_{\odot}$ which can also reproduce our data (see Section 5.2 and Figure \ref{GVTauS_spec_int}), so the allowed range of luminosities for GV Tau S likely spans a large range.

\section{Discussion}

\subsection{Gas vs. dust masses}

Until this point we have presented our models and results in terms of the mass of dust present in the system. Dust mass is constrained by our radiative transfer modeling, and is also the relevant quantity for understanding giant planet formation via core accretion \citep[e.g.][]{Lissauer1993}. However, disk masses are often quoted as the total of dust+gas mass. We therefore convert our dust masses into total masses using the common assumption that the gas-to-dust mass ratio is 100 times the total mass of our systems. With this assumption, the total mass in each of the GV Tau disks is 0.005 M$_{\odot}$. Throughout the remainder of the text we refer to the total mass rather than the dust mass.

\subsection{Comparison with previous works}

Prior to this work, several investigators have attempted to measure the disk masses of GV Tau N and S. \citet{Guilloteau2011} measured total disk masses of 0.0006 and 0.0005 M$_{\odot}$ for GV Tau N and S respectively, noting that the disks are likely optically thick and that these numbers are lower limits. Their results are compatible with our own as we find disk masses of 0.005 M$_{\odot}$ for both GV Tau N and S. \citet{Guilloteau2011} also measure disk radii of 17 and 10 AU for the disks in the system, again consistent with our constraints since neither work had a linear resolution of better than $\sim$50 AU. Furthermore, the smallest disk radius in our model grid was 30 AU, so it is possible that there is a similar or better quality fit model with a disk radius smaller than we considered in our modeling.

\begin{figure*}[h!]
\centering
\includegraphics[width=6in]{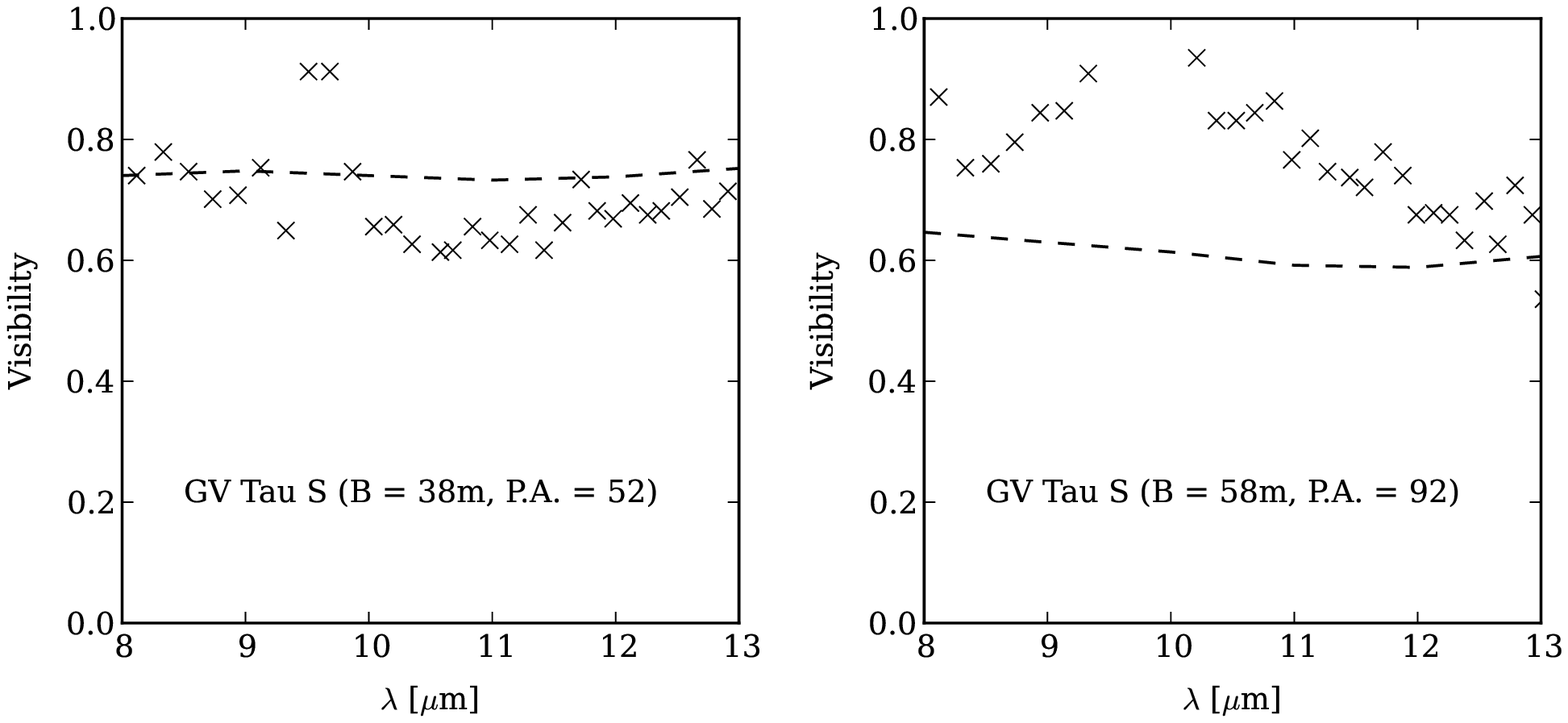}
\includegraphics[width=6in]{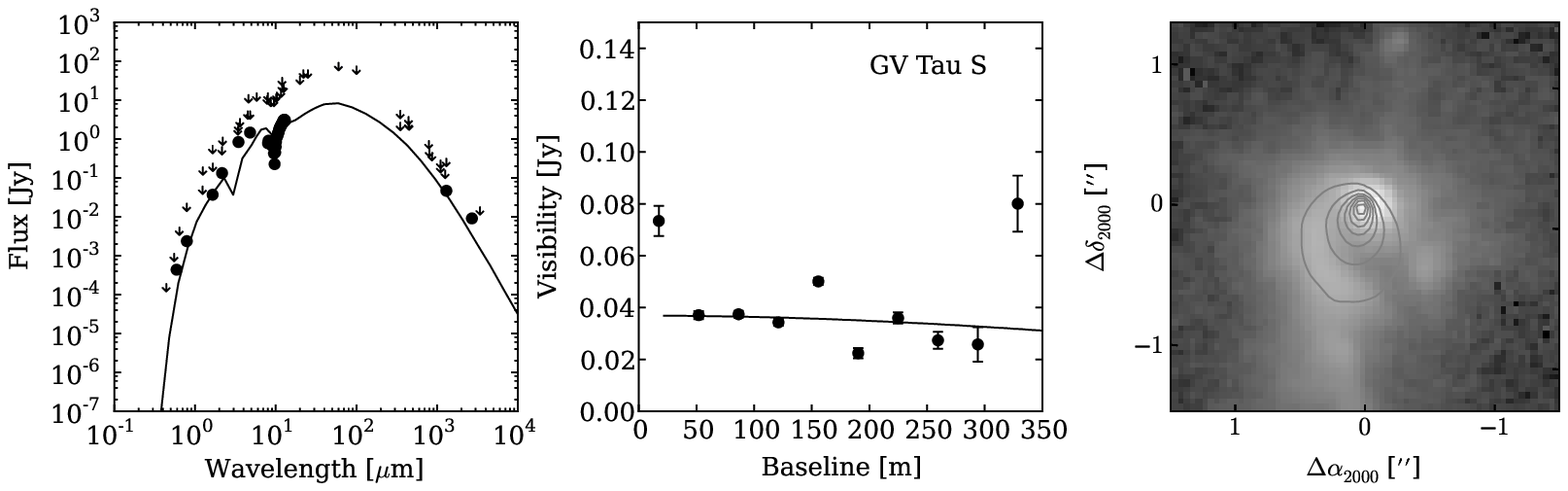}
\caption{Model which can plausibly reproduce the 8-13 $\mu$m visibilities for GV Tau S while also preserving the majority of our best fit model parameters. In the first row we show the 8-13 $\mu$m visibilities for GV Tau S from \citet{Roccatagliata2011} at two baselines, with the model visibilities shown as dashed lines. See Figure \ref{best_fit_opt} for more information about the panels in the second row. The parameters for this model are $L_{star} = 1.5$ L$_{\odot}$, $M_{disk} = 0.015$ M$_{\odot}$, $R_{disk} = 30$ AU, $h_0 = 0.01$ AU, $M_{env} = 0.002$ M$_{\odot}$, $R_{env} = 200$ AU, $f_{cav} = 0.2$, $\zeta = 0.7$, $R_{in} = 0.05$ AU, and $i = 55^{\circ}$. Although not perfect, the plot can plausibly reproduce all of the datasets while preserving the best fit parameters that we find from our modeling, within our estimated uncertainties.}
\label{GVTauS_spec_int}
\end{figure*}

There have been a number of studies which have made estimates of the inclination of the GV Tau S disk. \citet{Beck2010} detected spatially extended [Fe II] and Br$\gamma$ emission trailing from GV Tau S to the southwest, presumably tracing an outflow. They noted that the extent of the outflow is roughly consistent with an inclination of $60^{\circ}-70^{\circ}$, as suggested by \citet{Movsessian1999}. The same outflow is seen from GV Tau S at 3.6 cm by \citet{Reipurth2004}. Conversely, \citet{Roccatagliata2011} modeled 8-13 $\mu$m VLT visibilities with a two-blackbody model and found that GV Tau S is very close to face-on, with an inclination of $10^{\circ} \pm 5^{\circ}$. We find that our best fit model for GV Tau S, which includes physically-motivated complexity beyond the simple geometric model of \citet{Roccatagliata2011}, has a disk with an inclination of $55^{\circ}$. This matches the inclination found by \citet{Beck2010} but is decidedly different from that of \citet{Roccatagliata2011}. We are unable to find a model in our grid with an inclination consistent with \citet{Roccatagliata2011} that also fits all of the data well. We can, however, produce a model of GV Tau S that reproduces the 8-13 $\mu$m visibilities with an inclination of $55^{\circ}$ while maintaining the other parameters within their previously discussed uncertainties, as we demonstrate in Figure \ref{GVTauS_spec_int}, so we believe that these measurements are in fact consistent with a non-zero inclination.

\begin{figure*}[h!]
\centering
\includegraphics[width=6in]{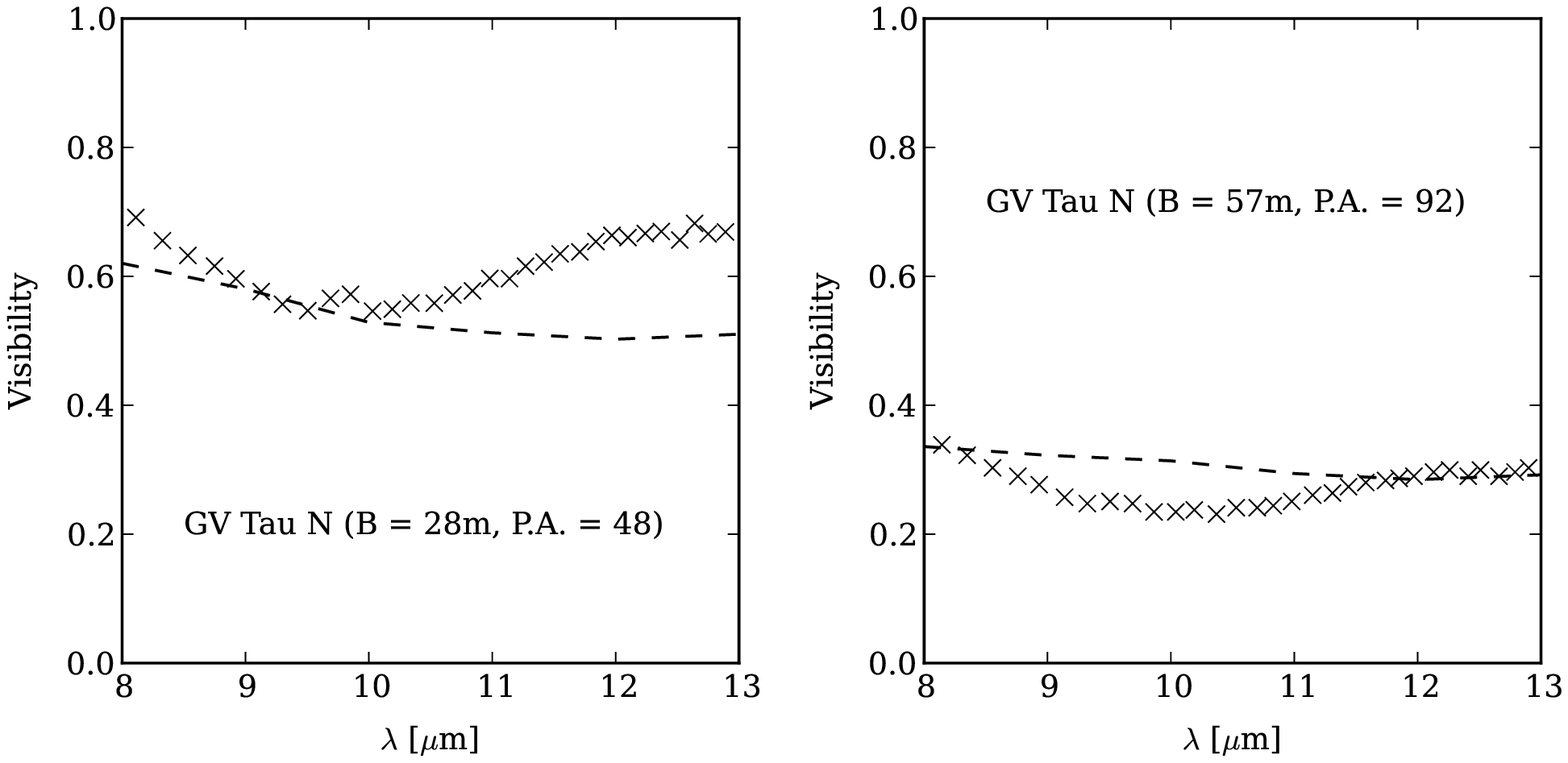}
\caption{8-13 $\mu$m visibilities measured by \citet{Roccatagliata2011} with the 8-13 $\mu$m visibilities for our best fit GV Tau N model overplotted as a dashed line. Each panel shows the visibilities at a different baseline. Our best fit model, which uses a very different inclination than is found by \citet{Roccatagliata2011} from the same data, can reproduce the best the data reasonably well, considering that the data were not included in our fitting. Indeed, the fit to the 10 um spectro-interferometry data is of comparable quality to the one presented in \citet{Roccatagliata2011}.}
\label{GVTauN_spec_int}
\end{figure*}

Little has been determined about the geometry of the disk of GV Tau N, although a number of studies have suggested that the disk is close to edge on based on the faintness of GV Tau N at short wavelengths. Furthermore, several investigators have reported the detection of warm HCN and/or C$_{2}$H$_{2}$ absorption in the disk of GV Tau N, which may suggest a higher inclination for the disk \citep{Gibb2007,Gibb2008,Doppmann2008,Fuente2012}. Indeed, \citet{Roccatagliata2011} measured an inclination for GV Tau N of $80^{\circ} \pm 10^{\circ}$ using VLT interferometry. Our work has demonstrated, however, that an edge-on disk cannot be invoked to reproduce the observed properties of GV Tau N. We are unable to find a model in our grid that can reproduce our datasets with an inclination consistent with the one found by \citet{Roccatagliata2011}. Our best fit model for GV Tau N suggests that the system has an inclination of $30^{\circ}$. Our best fit model can also plausibly reproduce the 8-13 $\mu$m visibilities modeled in \citet{Roccatagliata2011}, as we show in Figure \ref{GVTauN_spec_int}. We do not attempt to model the gas in the system so we cannot determine whether our best fit models are consistent with the detections of warm molecules towards GV Tau N.

A number of studies have suggested that the GV Tau system is surrounded by a flattened circumbinary envelope \citep{Menard1993,Koresko1999,Leinert2001}. As discussed earlier, we find that our simple double point source model for GV Tau is improved by adding a Gaussian source with a FWHM of $\sim$5". This Gaussian may represent emission from this circumbinary envelope. Because our interferometry data have limited sensitivity to extended emission, however, we cannot constrain the properties of such a circumbinary envelope.

Previous studies of near-infrared photometry have measured the luminosity of GV Tau S to be 1.8 L$_{\odot}$ \citep{White2004} and 3.3 L$_{\odot}$ \citep{Doppmann2005}, rougly consistent with our best fit models. Our models indicate a luminosity of 6 L$_{\odot}$ for GV Tau S, slightly higher than previous measurements, however we are also able to find acceptable model fits with luminosities as low as 1.5 L$_{\odot}$. As such, the previous measurements of the luminosity of GV Tau S fit nicely in the range allowed by our modeling.

\subsection{The Evolutionary State of GV Tau}
\label{GVTau_age}

GV Tau N and S are classified as Class I protostars, however previous investigators have estimated that the age of the system is $\sim3$ Myr \citep[e.g.][]{Doppmann2008}. That would mean that the protostars are more likely Class II pre-main sequence stars based on the ages of each stage as measured by counting statistics \citep[e.g.][]{Andre1994,Barsony1994}. If GV Tau were a Class II protostar, however, the highly obscured near infrared spectrum of GV Tau N would imply that the disk must be close to edge on, which is inconsistent with our modeling.

\citet{Doppmann2008} measure the age of the system by placing GV Tau N and S on an H-R diagram and comparing with pre-main sequence protostar tracks \citep{Siess2000}. They measure the temperature and surface gravity of each protstar by matching absorption line features in the near-infrared with stellar synthesis models and determine the mass by associating temperature with stellar mass. From there they use the mass and surface gravity to determine the stellar radius, and combine the radius and temperature to determine a luminosity. These measurements, however, are indirect, and are inconsistent with other measurements which use photometry and bolometric corrections to determine luminosity \citep{White2004,Doppmann2005}.

\begin{figure}[h!]
\centering
\includegraphics[width=3in]{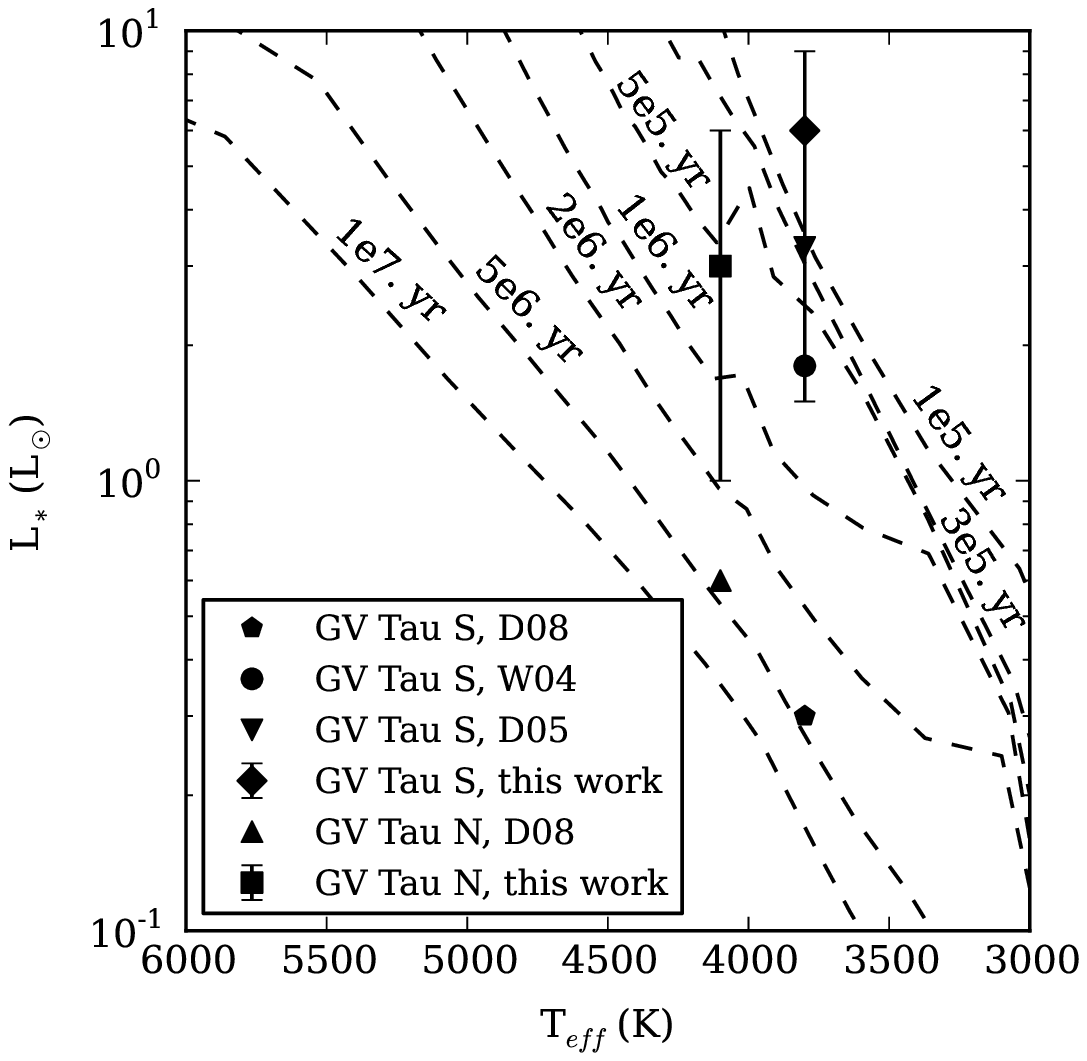}
\caption{Pre-main sequence tracks from \citet{Siess2000} as dashed lines with the temperature and luminosity measurements from \citet{White2004}, \citet{Doppmann2005}, \citet{Doppmann2008} and this work overplotted. The errorbars on our measurements represent the limited sampling of $L_{star}$ in our model grid rather than actual errors. Our luminosity measurements, as well as those of \citet{White2004} and \citet{Doppmann2005}, suggest much younger ages for the protostars than what \citet{Doppmann2008} measure. Our suggested age, of $\sim$ 0.5 Myr, is in better agreement with the ages for Class I protostars as estimated by counting statistics \citep[e.g.][]{Andre1994,Barsony1994}.}
\label{premsplot}
\end{figure}

If we use our luminosity constraints, or those of \citet{White2004} or \citet{Doppmann2005}, and the same evolutionary tracks to determine the age of the protostars, we find GV Tau has an age of a few hundred thousand years (see Figure \ref{premsplot}). This age provides a more consistent description of the system as a pair of Class I protostars, as suggested by the geometry of our best fit models, with an age of a few hundred thousand years. If we assume that the protostars are coeval, then we estimate an age of $\sim 0.5$ Myr for the system.

\subsection{Relation to the MMSN}

Disk mass is an important quantity for understanding the formation of planets. In order to form giant planets a protoplanetary disk must contain more than 0.01 M$_{\odot}$, and likely closer to 0.1 M$_{\odot}$, of material \citep{Weidenschilling1977,Desch2007}. Studies of the disks around Class II YSOs in Taurus and Orion (ages $\sim1-5$ Myr) have found that on average the disks around these stars do not contain enough material to form giant planets based on this criterion \citep{Andrews2005,Eisner2008, Andrews2013}. Observations at millimeter wavelengths, however, are only sensitive to dust grains smaller than a few millimeters. The insufficient mass present in the disks may be because dust grain growth in these disks hides the mass in larger bodies which are not traced by sub-millimeter observations.

\begin{figure*}[h!]
\centering
\includegraphics[width=6in]{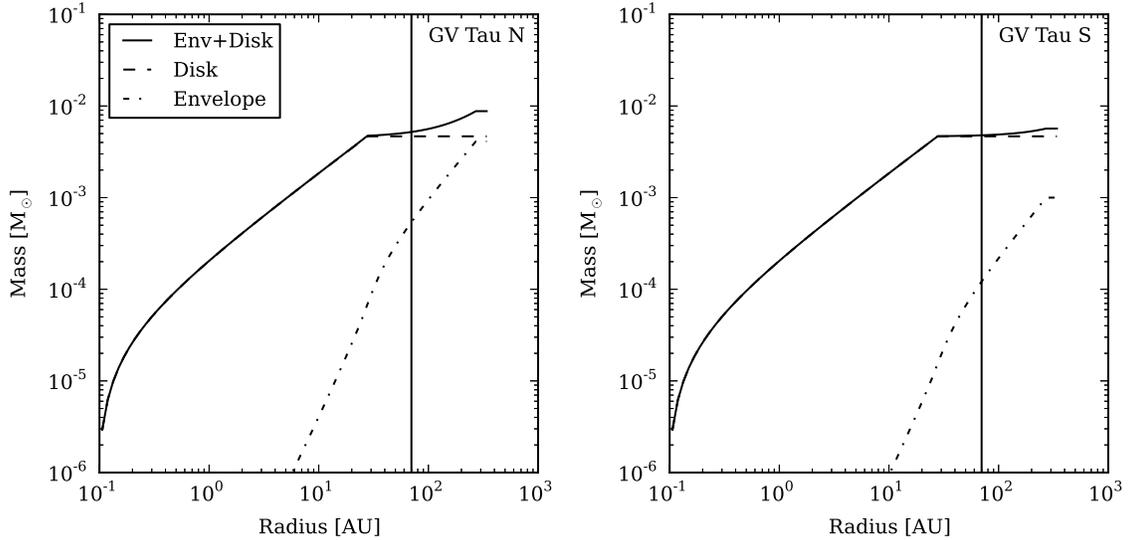}
\caption{Cumulative mass distribution for each component of the GV Tau system as a function of radius. For each figure we plot the contributions from the disk and envelope, as well as the combined distribution of the two. We also plot a vertical line at a radius of 70 AU (or a 140 AU diameter) to indicate the spatial resolution of our CARMA observations. While both protostars appear to have disks that are near or just shy of the 0.01 M$_{\odot}$ for forming giant plantets, all of the disk mass is within 30 AU, where giant planets are expected to form.}
\label{massplot}
\end{figure*}

We have estimated that the masses of the disks in the GV Tau system are $0.005$ M$_{\odot}$ each, which places both disks near the lower limit to the amount of matter needed to form giant planets \citep{Weidenschilling1977}. Unlike the other Class I disks measured in \citet{Eisner2012}, which were found to have a median disk mass within 100 AU of 0.007 M$_{\odot}$ and even less within 30 AU, the entirety of the disk mass in the GV Tau system is located within $\lesssim 30$ AU of the protostars. This is important because the Minimum Mass Solar Nebula (MMSN) is defined within 30 AU. The mass within 30 AU is then the proper mass to compare with the MMSN for determining potential for planet formation. While there may be just enough mass contained in the GV Tau circumstellar disks to form giant planets, the mass is located entirely within the regions of the disks where planets are actually formed. We plot the cumulative mass distribution for the GV Tau N and S best fit models as a function of radius in Figure \ref{massplot}.

If we include both components of GV Tau in the sample of Class I protostars in Taurus (ages $\sim0.1-1$ Myr) from \citet{Eisner2012} we find that the median mass of the Taurus Class I sample is 0.008 M$_{\odot}$. For comparison, the sample of Taurus Class II protostars (ages $\sim1-5$ Myr) from \citet{Andrews2013} has a median disk mass of 0.001 M$_{\odot}$. The sample of Orion Class II objects from \citet{Eisner2008} has a similar median disk mass. All ten of the disks in our Class I sample have a mass greater than or equal to the median mass of the sample from \citet{Andrews2013}. Fisher's exact test shows that the disks around our Class I sample are more massive than the Taurus and Orion Class II disks at the 99.8\% confidence level. The larger disk masses for Class I protostars likely reflects the fact that between the Class I and II stages some of the small dust particles in the disk have grown into larger bodies. Furthermore, the disk masses for both Class I and II protostars fall short of the minimum mass solar nebula, which may indicate that significant dust grain growth has already occurred by the time a protostar reaches the Class I stage.

We can also compare the Class I and Class II samples with the exoplanet sample to determine whether either distribution can reproduce the observed fraction of stars with giant planets. \citet{Cumming2008} determined that 18\% of stars have a giant planet within 20 AU, meaning that a minimum of 18\% of stars have giant planets. Conversely, the sample of Taurus and Orion Class II YSOs has 11\% of stars with disk masses greater than 0.01 M$_{\odot}$ and 0.6\% of stars with disk masses greater than 0.1 M$_{\odot}$ \citep{Andrews2005,Eisner2008,Andrews2013}. If we assume that the fraction of YSOs with disk masses sufficient to form giant planets is the same as the fraction of stars with giant planets, and we take 0.01 M$_{\odot}$ to be the threshold for forming giant planets, then the probability of randomly selecting Class II YSOs and reproducing the observed distribution is 0.02\%. If we take the threshold for forming planets to be 0.1 M$_{\odot}$ the probability becomes astronomically small. It would appear that the Taurus and Orion Class II disks cannot reproduce the observed fraction of stars with giant planets. For our Taurus Class I sample of 10 objects we find 2-7 objects that may have disks with masses greater than 0.01 M$_{\odot}$ but only 1 with a disk mass greater than 0.1 M$_{\odot}$. This would suggest that Class I protostars may have enough mass in their disks to form giant planets if the threshold is 0.01 M$_{\odot}$, but may not if the threshold is 0.1 M$_{\odot}$. Again, both of these comparisons neglect any disk mass in larger bodies that would not be traced well by observations.

\subsection{Stability of the Disks}

Previous studies have shown that the disk-to-star accretion rates measured for young stars are low compared with the time averaged envelope-to-disk infall rates \citep{Kenyon1987,White2004,Eisner2005}. One proposed solution to this discrepancy is that gravitational instabilities in the protostellar disk may lead to short bursts of gravitationally enhanced accretion in which material from the disk is rapidly accreted onto the central protostar. Such instabilities could be present in disks which are particularly massive, approximately one tenth the mass of the star, or dense.

\begin{figure*}
\centering
\includegraphics[width=6in]{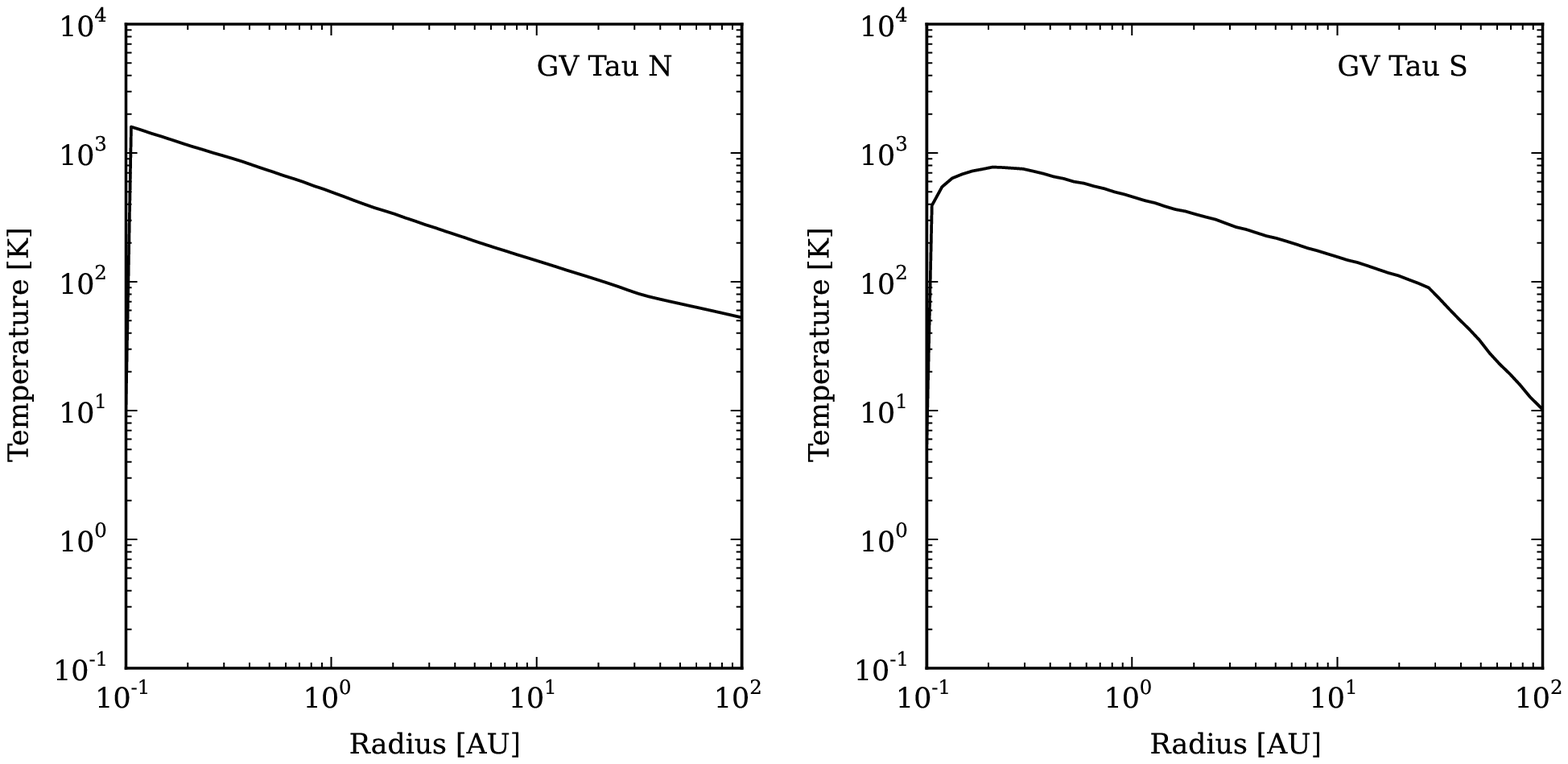}
\includegraphics[width=6in]{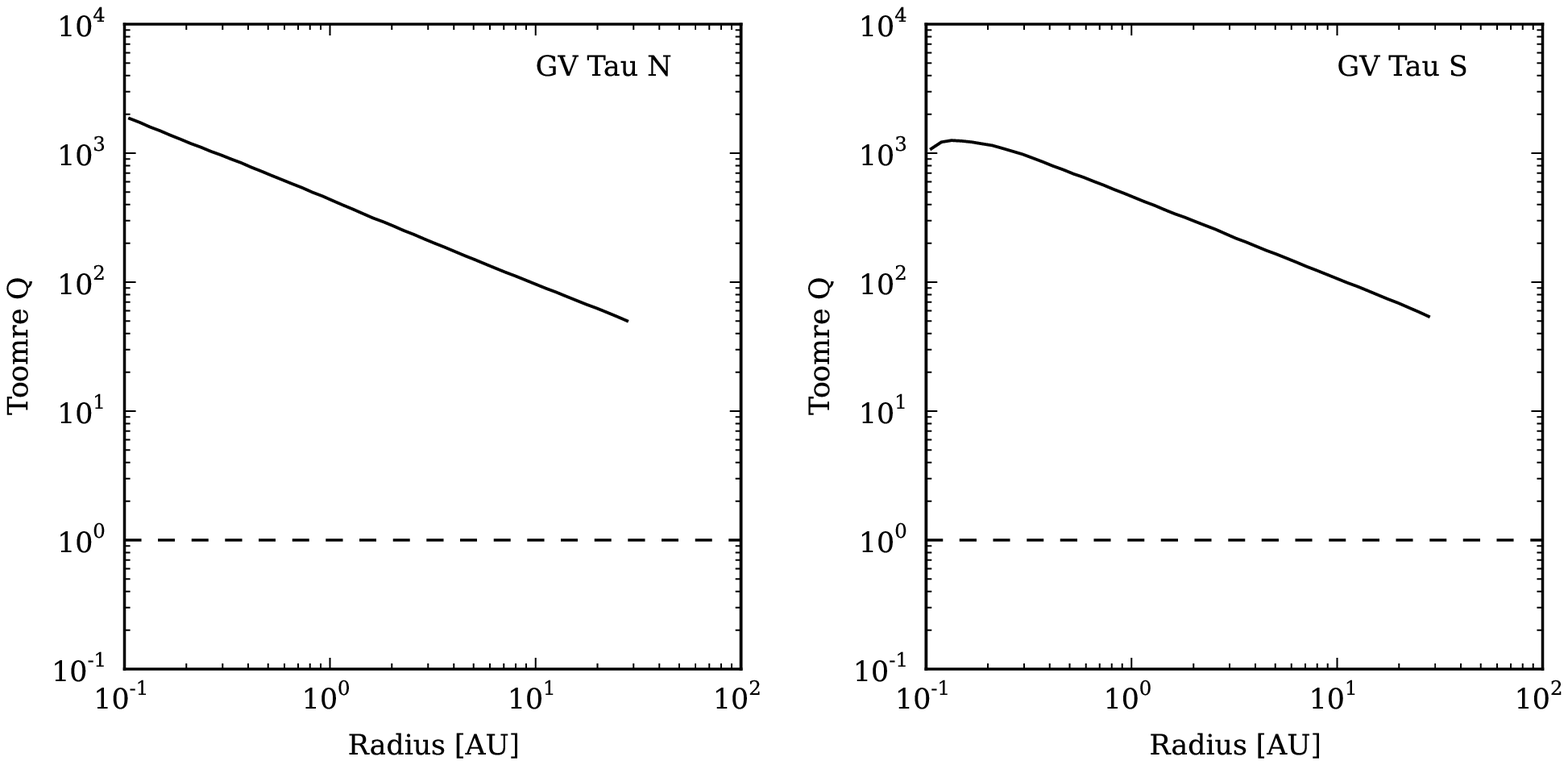}
\caption{({\it Top row}) Mean temperature of the disk as a function of radius for GV Tau N and S. ({\it Bottom row}) Toomre's Q as a function of radius for the disks of GV Tau N and S. The dashed line marks a value of Q=1. Values less than one imply that the disk is gravitationally unstable at that location, while values greater than one suggest that the disk is gravitationally stable. This shows that both disks are very stable. We do note that our millimeter-wave observations may not be sensitive to the entire mass of the disk, however we would have to be missing the majority of the total disk mass to make these disks unstable.}
\label{toomreplot}
\end{figure*}

Our study has suggested that the disks in the GV Tau system are quite small, on the order of 30 AU, but contain a significant amount of mass within those small disks. It might seem logical that these high density disks may be subject to gravitational instabilities. To calculate the stability of the disks we calculate Toomre's Q value as a function of radius for each disk. We assume a gas-to-dust ratio of 100 for this calculation. Values of Q $>1$ imply that the disk is stable against gravitational collapse, while values of Q $<1$ suggest that the disk may be susceptible to collapse under the force of gravity. We find that the best fit models for the GV Tau N and S disks are gravitationally stable throughout, with Q$>10$ at all radii. Our millimeter-wave observations are likely not sensitive to all of the mass in the disk, however we would have to be missing a majority of the mass to make these disks unstable. We plot the value of Toomre's Q as well as the mean disk temperature as a function of radius for both disks in Figure \ref{toomreplot}.

\subsection{Formation Mechanism}

There have been a number of proposed mechanisms that may lead to the formation of a binary star. One potential route occurs when a molecular cloud core that has begun to collapse to form a protostar fragments into multiple cores, each of which in turn collapse to form individual stars in a binary system \citep[e.g.][]{Boss1979,Bate1997}. Alternatively, a binary star system could be formed when the protostellar disk surrounding a young star becomes gravitationally unstable and collapses to form a second star in the system \citep[e.g.][]{Bonnell1994,Bonnell1994a,Bonnell1994b,Burkert1997}.

Both proposed theories make predictions about the geometry of the resulting binary system that can be used to explore how a binary system formed. A binary system formed by a gravitational instability in the disk around the primary star is expected to have protoplanetary disks that are aligned. Disks that formed, however, by the fragmentation of a collapsing molecular cloud core can be misaligned \citep{Bate2000}. Interactions with passing objects or the accretion of a small amount of material with a different angular momentum near the end of the accretion phase can also cause misalignment \citep{Bate2000}. Conversely, tidal interactions can act to align disks, as well.

\citet{Roccatagliata2011} measured the inclination of the disks in the GV Tau system to be $10^{\circ}$ and $80^{\circ}$ and claimed that this misalignment is evidence that the system formed as the result of molecular cloud core fragmentation. Our best fit model for each component finds inclinations of $30^{\circ}$ and $55^{\circ}$, although each of these estimates may be able to vary by $20^{\circ}$. The mutual inclination of the disks is then close to $25^{\circ}$, but may vary by $30^{\circ}$. These results suggest that the mutual inclination of the disks in the GV Tau system may not be as high as \citet{Roccatagliata2011} found, and the disks may even be aligned. Thus we cannot distinguish between formation scenarios.

\subsection{Future Work}

While we are able to constrain the mass in the GV Tau protoplanetary disks with our current data, we have left a number of other parameters somewhat unconstrained. Higher spatial resolution millimeter observations of the binary can resolve the protoplanetary disks and more accurately measure the disk radii, inclinations, and position angles. If the disks have radii of 30 AU or smaller, as we have suggested, then a resolution of $< 0.4"$ will allow us to resolve the disks and make these measurements. Modern interferometers, such as ALMA or the VLA, can provide high enough spatial resolution ($\sim 0.05"$ for the VLA) to resolve these disks and determine these parameters without ambiguity. 

It may also be important to add observations of GV Tau with an interferometer in a more compact configuration. Such a configuration would be significantly more sensitive to the faint extended emission from the proposed circumbinary envelope. Our observations with the CARMA C-array do not include baselines shorter than $\sim 20$m. As such we are not very sensitive to spatial scales larger than about 1000 AU. Compact array observations would be capable of detecting the signal from this extended envelope around GV Tau and could be important important for breaking any degeneracy between the envelope mass and radius.

Another significant source of uncertainty in our measurements comes from the opacity of the dust assumed for our model. In this study we were unable to constrain the dust grain properties in the system. One way to better constrain the opacity for the dust in the protoplanetary disks is with additional millimeter-wave observations of the system. Millimeter fluxes of dust roughly follow a power law, $F_{\nu} \propto \nu^{2+\beta}$, where $\beta$ is related to the optical properties of the dust, with $\beta \sim 2$ corresponding to small grains and $\beta \sim 0$ relating to larger dust grains. Multiple millimeter wavelength observations can thus help to constrain the dust optical properties. There is some evidence that $\beta \approx 0$ in GV Tau N, if the 3.6cm emission seen by \citet{Reipurth2004} is from dust emission, so GV Tau N may be a particularly interesting candidate for this sort of study.

\section{Conclusion}

We have used detailed radiative transfer modeling to create synthetic model protostars to match to CARMA millimeter visibilities, HST near-infrared scattered light imaging, and broadband SEDs in order to constrain the masses of the disks around the protostars in the binary YSO system GV Tau. We find that the best fit model disks around GV Tau N and S each have gas+dust masses of 0.005 M$_{\odot}$ and disk radii $< 30$ AU, and that the age of the system is $\sim$ 0.5 Myr. These estimates place both components near the lower end of the Minimum Mass Solar Nebula, meaning they may have just enough mass to form giant planets. We also find that both disks are gravitationally stable throughout, unless our millimeter-wave observations are missing the majority of the disk mass. Furthermore, we find that the disks of GV Tau N and S are inclined at $30^{\circ}$ and $55^{\circ}$ respectively, consistent with some previous studies of the system \citep{Movsessian1999,Beck2010}, but inconsistent with a recent study by \citet{Roccatagliata2011}. We have shown, however, that we can plausibly reproduce the 8-13 $\mu$m visibilities from \citet{Roccatagliata2011} with our best fit model for GV Tau N and a modified version of our best fit model for GV Tau S which preserves the inclination of our best fit model.

When we include both protostars in the GV Tau system with the Class I protostars modeled by \citet{Eisner2012} we find that the sample of 10 Class I protostars has a median disk mass of 0.008 - 0.01 M$_{\odot}$. All of the disks in our Class I sample are more massive than the median of the Class II sample of disks (of 0.001 M$_{\odot}$). These numbers suggest that, on average, the circumstellar disks of Class I protostars are more massive than those of the more evolved Class II protostars. This likely indicates that between the two stages some of the smaller dust grains in the disks have grown into larger bodies. For both samples, however, the median masses fall below the minimum mass solar nebula \citep{Weidenschilling1977,Desch2007}, and may not be able to reproduce the observed frequency of giant planets. It may be the case that significant dust grain processing has already occurred by the Class I stage, and it may be necessary to explore even younger disks to determine the initial mass budget for planet formation.

\acknowledgements

The authors would like to gratefully acknowledge Thomas Robitaille and Kees Dullemond for making their codes, Hyperion and RADMC-3D, publicly available. This material is based upon work supported by the National Science Foundation Graduate Research Fellowship under Grant No. 2012115762. JAE acknowledges support from an Alfred P. Sloan Foundation fellowship. Support for CARMA construction was derived from the Gordon and Betty Moore Foundation, the Kenneth T. and Eileen L. Norris Foundation, the James S. McDonnell Foundation, the Associates of the California Institute of Technology, the University of Chicago, the states of California, Illinois, and Maryland, and the National Science Foundation. Ongoing CARMA development and operations are supported by the National Science Foundation under a cooperative agreement, and by the CARMA partner universities.

\bibliographystyle{apj}
\bibliography{ms}

\end{document}